\documentclass[11 pt]{article}

\usepackage{xfrac}
\usepackage{xcolor}
\usepackage[toc,page]{appendix}

\usepackage{amsmath}
\usepackage{amsfonts}
\usepackage{slashed}
\usepackage{caption}
\usepackage{float}
\usepackage{physics}
\usepackage[capitalize]{cleveref}

\mathchardef\mhyphen="2D

\usepackage{graphicx}
\graphicspath{ {./images/} }

\newcommand{\rk}{\mathrm{k}}        

\begin{document}

\vspace{20pt}
\begin{center}
{\Large\bf Generalised spectral dimensions in non-perturbative quantum gravity}
\vspace{15pt}

{\large M. Reitz$^a$, D. Németh$^a$, D. Rajbhandari$^a$, A. G\"orlich$^{a,b}$, J. Gizbert-Studnicki$^{a,b}$}
\vspace{15pt}

$^{a}${\sl Institute of Theoretical Physics, Jagiellonian University,\\
Łojasiewicza 11, Kraków, PL 30-348, Poland.}\\
$^{b}${\sl Mark Kac Center for Complex Systems Research, Jagiellonian University,
Łojasiewicza 11, Kraków, PL 30-348, Poland.}
\vspace{15pt}

email:  {\sl marcus.reitz@uj.edu.pl}
\vspace{40pt}

\begin{abstract}
The seemingly universal phenomenon of scale-dependent effective dimensions in non-perturbative theories of quantum gravity has been shown to be a potential source of quantum gravity phenomenology. The scale-dependent effective dimension from quantum gravity has only been considered for scalar fields. It is however possible that the non-manifold like structures, that are expected to appear near the Planck scale, have an effective dimension that depends on the type of field under consideration. To investigate this question, we have studied the spectral dimension associated to the Laplace-Beltrami operator generalised to $k$-form fields on spatial slices of the non-perturbative model of quantum gravity known as Causal Dynamical Triangulations. We have found that the two-form, tensor and dual scalar spectral dimensions exhibit a flow between two scales at which an effective dimension appears. However, the one-form and vector spectral dimensions show only a single effective dimension. The fact that the one-form and vector spectral dimension do not show a flow of the effective dimension can potentially be related to the absence of a dispersion relation for the electromagnetic field, but dynamically generated instead of as an assumption.
\end{abstract}
\end{center}
\newpage
\tableofcontents
\newpage

\section{Introduction}

An important tasks of quantum gravity research is to find observables that show possible deviations from classical gravity. Especially in view of the recent availability of gravitational wave data, the search for new potentially observable signatures of quantum gravity has a high priority. An example of deviation from classical gravity coming from quantum gravity is the possibility of a scale-dependent effective dimension of spacetime. This running of the spectral dimension has been discovered to be present in many different theories of quantum gravity \cite{Carlip_2017, Carlip,  Lifshitz, Trzesniewski}.

The interpretation of this seemingly universal quantum gravity phenomenon is that spacetime is not well described by a manifold of a fixed dimension near the Planck scale. Although it is not entirely clear how a running spectral dimension would manifest itself in real-world observations, there have been attempts to formulate model-independent signatures of the running spectral dimension in gravitational wave physics \cite{Calcagni1, Calcagni2, Calcagni3}. While showing that there is a potential for an observable effect in gravitational wave observations, the analysis of signatures of the running spectral dimension in gravitational wave physics relies on a number of assumptions. It is assumed that only the scalar mode of the gravitational waves is sensitive to the running spectral dimension and that other modes are unaffected. It is furthermore assumed that the electromagnetic field, which is a vector field, is also unaffected.

The spectral dimension is a quantity related to the spectrum of the Laplacian, which is a second-order operator on the space of scalar functions. A generalisation of the Laplace operator is the Laplace-Beltrami\footnote{Sometimes $\Delta^{(k)}$ is called the Laplace–de Rham operator.} operator $\Delta^{(k)}$, defined as a second order differential operator on the space of $k$-forms. The Laplace-Beltrami operator $\Delta^{(k)}$ also defines a generalised spectral dimension $D^{(k)}$. For a differentiable Riemannian manifold $\mathcal{M}$, the spectral dimension is identical for all $k$. However, because a single differentiable manifold is most likely not the correct description of spacetime when quantum gravitational effects are relevant, it is an important question whether the generalised spectral dimension $D^{(k)}$ depends on $k$ in quantum gravity. The Laplace-Beltrami operator $\Delta^{(k)}$ is also related to propagation of $k$-form fields. A $k$-dependent spectral dimension $D^{(k)}$ can therefore signal that different fields and gravitational wave modes have a different scale dependence. As far as we are aware\footnote{A framework to investigate $\Delta^{(k)}$ for all $k$ in discrete models of quantum gravity was discussed in \cite{Oriti}, but only the dual scalar case was considered in detail.}, the only cases that have been studied in quantum gravity are those associated to scalar and dual scalar fields. The goal of this work is to investigate whether the generalised spectral dimension $D^{(k)}$ depends on $k$ in quantum gravity and whether it is scale dependent for all $k$. This work is a step towards explicitly testing the assumptions made in deriving signatures of the running spectral dimension in gravitational wave physics and improving our knowledge on how quantum gravity can potentially manifest itself.

The running of the spectral dimension was first discovered in Causal Dynamical Triangulations (CDT) \cite{SpecDimCDT1}, a non-perturbative approach to quantum gravity that is defined in terms of a lattice regularisation of the gravitational path-integral \cite{CDTReview2012,CDTReview2019, CDTReview2021}. One of the advantages of CDT is that it is particularly suited for numerical simulations. Therefore, it is possible to calculate observables that are not easily accessible in other frameworks. CDT is defined in terms of piecewise flat manifolds or triangulations $\mathcal{T}$, a specific type of simplicial complexes. Using well known tools from the study of simplicial complexes \cite{ComLap}, it is relatively straightforward to define the Laplace-Beltrami operator $\Delta^{(k)}$ in CDT. We can therefore use the numerical tools of CDT to study the $D^{(k)}$ in a non-perturbative formulation of quantum gravity. In section \ref{CDT} we will review those aspects of CDT that are important for the remainder of the text. Section \ref{Continuum} contains a discussion of the Laplace-Beltrami operator and the associated generalised spectral dimension. Section \ref{Discretuum} explains how the Laplace-Beltrami operator and the generalised spectral dimension can be generalised to simplicial complexes. Section \ref{Results} contains the measurements of the generalised spectral dimension in CDT. Finally, we will summarise our results in section \ref{Discussion} and give an interpretation in terms of quantum gravity phenomenology.

\section{Overview of CDT} \label{CDT}
\begin{figure}[ht!]
    \centering
    \includegraphics[width=0.8\textwidth]{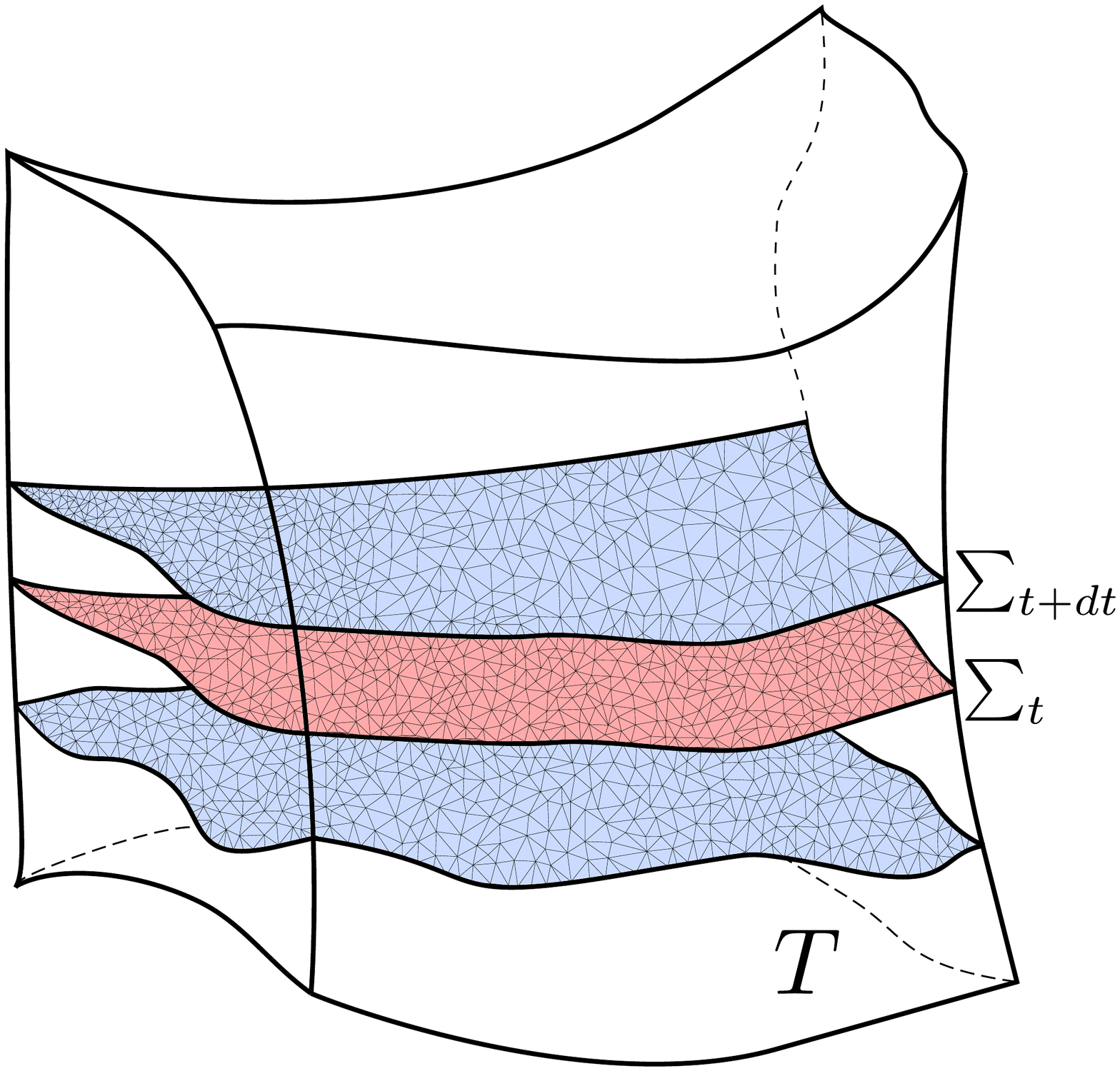}
    \caption{The figure shows a CDT spacetime $T$ foliated by a family of space-like hypersurfaces $(\sum_{t})_{t \in \mathbb{Z}}$}
    \label{fig:cdt-spacetime}
\end{figure}

\begin{figure}[h]
    \centering
    \includegraphics[width = 0.8\textwidth]{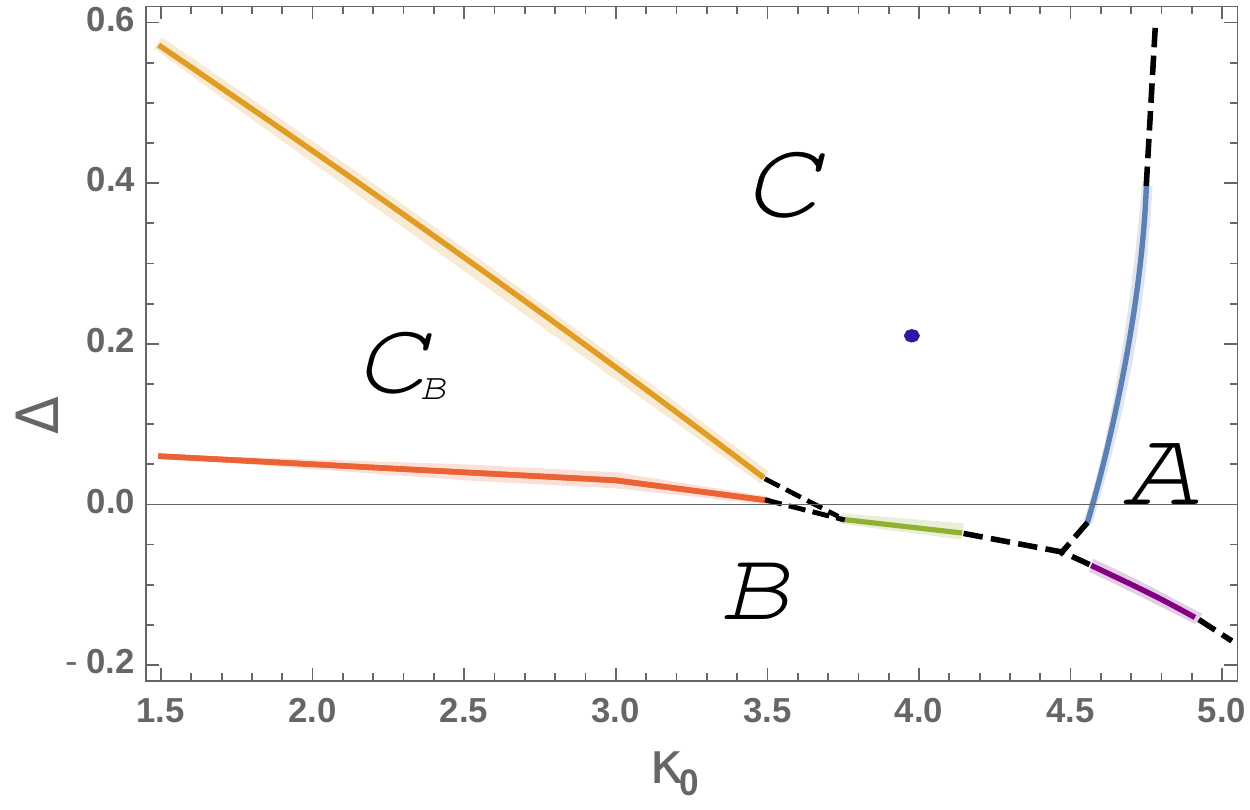}
    \caption{The phase diagram of CDT with its four phases. Phase $A$ is the branched polymer phase, $B$ is the collapsed phase, $C_b$ the bifurcation and $C$ the de Sitter phase. The purple dot at coordinate $(\kappa_0,\Delta) = (4.0, 0.2)$ represents the location where the configurations were sampled.}
    \label{fig:my_label}
\end{figure}

CDT is a non-perturbative model of quantum gravity \cite{CDTReview2012, CDTReview2019, CDTReview2021}, based on a lattice regularisation of the formal gravitational path integral $Z$,
\begin{equation}
    Z = \int \mathcal{D}[g_{\mu\nu}]e^{iS_{EH}[g_{\mu\nu}]}.
\end{equation}
The regularisation procedure consists of a discretisation of spacetime and replacing the Einstein-Hilbert action $S_{EH}$ with the (Lorentzian) Regge action $S^{(L)}_{R}$ \cite{Regge:1961px}. Formal integration over metric manifolds is replaced by a sum over discrete spacetimes $T \in \mathcal{T}^{(L)}_{CDT}$,
\begin{equation}
    Z = \int \mathcal{D}[g_{\mu\nu}]e^{iS_{EH}[g_{\mu\nu}]} \longrightarrow \sum_{T\in \mathcal{T}^{(L)}_{CDT}}
    \frac{1}{C_T}e^{iS_{R}^{(L)}[\mathcal{T}]}.
\end{equation}
The factor $C_T$ is the order of the automorphism group of $T$. The set of CDT geometries $\mathcal{T}^{(L)}_{CDT}$ consists of spacetimes that are $n$-dimensional simplicial Lorentzian manifolds\footnote{These simplicial manifolds, or triangulations, consist of a few types of identical $n$-dimensional Lorentzian simplices glued together along $(n-1)$-dimensional faces.}, determined by a lattice spacing $a$, that admit a globally hyperbolic foliation. Each leaf of the foliation is a $(n-1)$-dimensional simplicial manifold (see figure \ref{fig:cdt-spacetime}) of fixed topology and is labelled by a discrete time step $t$. The foliation ensures a causal evolution of spatial hypersurfaces. The lattice regularisation can in principle be removed by a simultaneous limit of the number of simplices $N \rightarrow \infty$ and of the lattice spacing $a \rightarrow 0$.
The foliation also makes a well-defined Wick rotation possible. The Wick rotation transforms the Lorentzian action $S^{(L)}_R$ into the Euclidean action $S_R$ and the set of Lorentzian triangulations $\mathcal{T}^{(L)}_{CDT}$ into the set of Euclidean triangulations $\mathcal{T}_{CDT}$,
\begin{equation}
    Z = \sum_{T \in \mathcal{T}_{CDT}} \frac{1}{C_T}e^{-S_{R}[T]},
\end{equation}
allowing for a statistical treatment of the model. In 3+1 dimensions, after the Wick rotation the Euclidean Regge action $S_{R}$ is given by
\begin{equation}
S_{R} = -(\kappa_0+6\Delta)N_0 + \kappa_4(N_{41}+N_{32}) + \Delta N_{41}. 
\label{CDTAction}
\end{equation}
$N_{0}$, $N_{41}$ and $N_{32}$ are the global parameters related to the total number of vertices, $\{4,1\}$-type of simplices and $\{3,2\}$-type of simplices, respectively. The number $i$ for an $\{i,j\}$-type of simplex denotes the number of vertices of a simplex at time $t$ and the number $j$ denotes the number of vertices at time $t\pm 1$. The bare couplings that appear in $S_R$ are the (inverse) bare gravitational constant $\kappa_0$, the bare cosmological constant $\kappa_4$ and the asymmetry parameter $\Delta$. The coupling $\Delta$ determines the difference in proper length between space-like and time-like links of the simplices. The expectation value of an observable $\mathcal{O}$ is given by,
\begin{equation}
\langle \mathcal{O} \rangle = \frac{1}{{Z}}\sum_{T}\frac{1}{{C}_{T}}\mathcal{O}(T)e^{-S_{R}[T]}.
\label{ExVal}
\end{equation}
In the absence of an exact calculation of $\langle \mathcal{O} \rangle$, one can sample the CDT configuration space with Monte Carlo simulations and approximate $\langle \mathcal{O} \rangle$ by averaging $\mathcal{O}$ over the resulting ensemble of geometries generated with the Boltzmann weight $e^{-S_R[T]}$. The Monte Carlo simulations are performed for a range of target volumes $\bar{N}_{41}$, controlled by the additional volume fixing term $S_{fix}=\epsilon (N_{41}-\bar N_{41})^2$ added to the Regge action\footnote{In order to force the volume $N_{41}$ of the triangulation to fluctuate around the target $\bar N_{41}$ it is also necessary to tune the cosmological coupling constant $\kappa_4\to \kappa_4^{c}(\kappa_0, \Delta,\bar N_{41})$. This way one trades the $\kappa_4$ coupling constant with the target volume $\bar N_{41}$.}. Subsequently, the average of $\mathcal{O}$ can be extrapolated to the infinite-volume limit using standard finite-volume scaling analysis. The theory space of CDT can be investigated by varying the bare coupling constants. The phase diagram of CDT (see Fig. \ref{fig:my_label}) has a rich structure \cite{thephasestruc}.

Out of four phases with distinct geometric properties, the most interesting one is phase $C$, also called the de Sitter phase, where on large scales (compared to the size of the triangulations) one observes an emerging four-dimensional geometry \cite{SpecDimCDT2}. In was shown  that in phase $C$ the volume profile obtained by integrating out all degrees of freedom except the scale factor is consistent with classical solutions of General Relativity and quantum fluctuations of the scale factor are well described by the  minisuperspace-type effective action \cite{nonpertquantumdes,impacttop}. This result required an introduction of appropriate observables which can be measured and averaged over in the Monte Carlo simulations. One such observable was the (dual) scalar spectral dimension which, as already mentioned, showed a non-trivial scale dependence, changing from 4 in large scales to approximately 2 in short scales \cite{SpecDimCDT1}. This result suggests that even though the large scale behaviour of the quantum geometry looks (semi-)classical, the short scale features can be very far from any behaviour of classical geometries and most likely cannot be described in terms of smooth differentiable manifolds. It was indeed shown that the leaves of the CDT foliation (the spatial slices) have very non-trivial fractal geometry \cite{SpecDimCDT2,AMBJORN2010420}. The four-dimensional geometry is also non-smooth on short scales \cite{SpecDimCDT2,Zbyszek2021}. We therefore expect that by performing numerical simulations in phase $C$ and studying observables sensitive to  short scales one can investigate features of the UV limit of quantum gravity, while  observables sensitive to large scales will at the same time show features of the IR limit. The scale-dependent generalized spectral dimensions $D^{(k)}$ can therefore be used to simultaneously investigate both regimes. 

\section{Tensor Diffusion}
\subsection{Continuum tensor diffusion}
\label{Continuum}
The heat equation governs diffusions and random walks on a general $n$-dimensional Riemannian manifold $(\mathcal{M}, g)$. We will only consider the case where $\mathcal{M}$ is closed. The heat equation,
\begin{equation}
\frac{\partial}{\partial \tau}K(x,x_0;\tau) = -\Delta K(x,x_0;\tau),
\end{equation}
describes the evolution of the heat kernel $K(x, x_0; \tau)$, which represents the probability density at position $x\in \mathcal{M}$ of finding a particle after a random walk over a diffusion time $\tau$. The initial condition for a particle which originated at position $x_0$ at $\tau = 0$ is given by a delta function,
\begin{equation}
K(x, x_0; \tau = 0) = \delta^n(x - x_0).
\label{delta}
\end{equation}
The total probability is preserved over time,
\begin{equation}
\int dx \sqrt{g} K(x,x_0;\tau) = 1.
\end{equation}
The average return probability $P(\tau)$ of the random walk returning to the origin after time $\tau$, averaged over the manifold $\mathcal{M}$, can also be defined by the heat kernel,
\begin{equation}
P(\tau) = \frac{1}{V}\int dx \sqrt{g} K(x,x;\tau),
\end{equation}
where $V = \int dx \sqrt{g}$ is the volume of $\mathcal{M}$. The return probability can be expressed exactly for all $\tau$ in terms of the eigenvalues $\lambda_i$ of the Laplacian $\Delta$,
\begin{equation}
P(\tau) = \frac{1}{V}\sum_{i=0} e^{-\tau\lambda_i},
\label{RetEig}
\end{equation}
which is identical to the trace of the heat kernel. The asymptotic expansion of $P(\tau)$ for small $\tau$ is known \cite{Craioveanu},
\begin{equation}
P(\tau) \sim \tau ^{-\frac{n}{2}}\sum_{i=0} A_i\tau^i,
\label{Ret}
\end{equation}
where $A_i$ are generally complicated integrated functions of the metric $g$.
In equation \eqref{RetEig} we assume that the spectrum is discrete, which is valid if $\mathcal{M}$ is a closed manifold. For an open manifold, the sum can be replaced by an integration.

Analogous expressions to Eqs. \eqref{RetEig} and \eqref{Ret} can be found for a generalised heat equation based on the Laplace-Beltrami operator $\Delta^{(k)}= d\delta+\delta d$, defined on $k$-forms. Here $d$ is the exterior derivative and $\delta=(-1)^{n(k-1)+1}\star \! d \star$ is the co-differential. The $k$-form heat equation is,
\begin{equation}
\frac{\partial}{\partial \tau}\omega(x,x_0;\tau) = -\Delta^{(k)} \omega(x,x_0;\tau),
\label{kHE}
\end{equation}
where $\omega(x,x_0;\tau)$ is a $k$-form valued heat kernel. The analogue of the average return probability $P^{(k)}(\tau)$ for $k$-forms, $0\leq k\leq n$, is identical to the trace of the $k$-form heat kernel, as was the case for the average return probability in equation \eqref{RetEig}. It can also be expressed in terms of the eigenvalues $\lambda^{(k)}_i$ of $\Delta^{(k)}$,
\begin{equation}
P^{(k)}(\tau) = \frac{1}{V}\sum_{i=0} e^{-\tau \lambda^{(k)}_i}.
\label{RetEigk}
\end{equation}
The asymptotic expansion of the function $P^{(k)}(\tau)$ has also essentially the same form,
\begin{equation}
P^{(k)}(\tau)\sim \tau^{-\frac{n}{2}}\sum_{i=0} B_i(\Delta^{(k)})\tau^i, 
\end{equation}
where again the $B_i(\Delta^{(k)})$ are in general complicated functions of the metric $g$.

The dimension of the differentiable manifold $n$ can be extracted from the asymptotic expansion,
\begin{equation}
n = - 2 \lim_{\tau \rightarrow 0} \frac{d\log P^{(k)}_\tau }{d\log \tau}.
\label{dim}
\end{equation}
The expression in equation \eqref{dim} can be generalised by dropping the limit $\tau \rightarrow 0$. The resulting object is the running spectral dimension $D^{(k)}(\tau)$, which is a function of $\tau$,
\begin{equation}
D^{(k)}(\tau) = -2\frac{d\log P^{(k)}_\tau }{d\log \tau}.
\label{specdim}
\end{equation}
For later use, we will comment here on the effect of a global rescaling of the Laplacian $\Delta^{(k)}$ by a factor $b$, $\Delta^{(k)} \rightarrow b \Delta^{(k)}$. The rescaling of $\Delta^{(k)}$ is equivalent to a rescaling of the diffusion time $\tau$, which will leave the spectral dimension invariant. This equivalence is apparent from the change of the return probability under the rescaling,
\begin{equation}
    P(\tau) \rightarrow P'(\tau) =\frac{1}{V} \sum_i e^{-\tau b \lambda_i} = P(\tau'), \quad  \tau' = b\tau.
\label{RetScale}
\end{equation}
From Eqs. \eqref{RetEigk} and \eqref{specdim} we can see that the running spectral dimenision $D^{(k)}(\tau)$ goes to zero for large $\tau$ if $\Delta^{(k)}$ has at least one zero eigenvalue\footnote{The number of zero-valued eigenvalues of $\Delta^{(k)}$ is a topological property of $\mathcal{M}$ and is given by the Betti numbers $b_i, \ i \in \{0,...,n\}$, of $\mathcal{M}$. If $\mathcal{M}$ consists of a single connected component, the Betti numbers $b_0$ and $b_n$ are equal to $1$. The Betti numbers also adhere to the symmetry relation $b_i=b_{n-i}$.}. At diffusion time $\tau$, the eigenvalues $\lambda_i\leq \tau^{-1}$ dominate the behaviour of the return probability $P^{(k)}(\tau)$. The smallest non-zero eigenvalue $\lambda_1$ is bounded from below, $\lambda_1 > r^{-2}$, where $r\sim V^{1/n}$ is the diameter of $\mathcal{M}$ \cite{He}. We conclude that, for timescales,
\begin{equation}
\tau > V^{2/n},
\label{GapDom}
\end{equation}
the zero mode of $\Delta^{(k)}$ becomes dominant and the diffusion process is strongly affected by the finite size of $\mathcal{M}$.

The scalar case, corresponding to $k=0$, has been extensively studied\footnote{Depending on the theory of quantum gravity, often the dual scalar case is studied corresponding to $k=n$.} in various approaches to quantum gravity \cite{Carlip, Carlip_2017, Lifshitz, Trzesniewski} and is generally interpreted as a scale-dependent effective dimension. The goal of this work is to generalise the studies of scale-dependent effective dimensions in quantum gravity to general $k$-forms. In the next section we will therefore discuss the discrete analogue of the running spectral dimension $D^{(k)}(\tau)$ for $k$-forms.

\subsection{Discrete tensor diffusion}
\label{Discretuum}

The Laplace-Beltrami operator $\Delta^{(k)}$ can be generalised to an equilateral simplicial complex $T$ \cite{MCA2}. To make a clear distinction we will write $L^{(k)}$ for the discrete analogue of $\Delta^{(k)}$. The operator $L^{(k)}$ is defined\footnote{See \cite{Desbrun} for an introduction to the topic of discrete differential forms.} on the space of discrete $k$-form fields $\Omega^k(T)$, which are associated to the $k$-simplices $\sigma^k$ that are contained in $T$ and can be represented as a real symmetric matrix of rank $N_k$, the number of $k$-simplices in $T$. The operator $L^{(k)}$ can be decomposed in two terms, just as $\Delta^{(k)}$, and is given by,
\begin{equation}
L^{(k)} = L^{(k)}_++L^{(k)}_-, \quad L^{(k)}_+=(I^{k+1}_k)^T I^{k+1}_k, \quad L^{(k)}_-=I^k_{k-1} (I^{k}_{k-1})^T,
\label{DiscLap}
\end{equation}
where $I^{k+1}_k$ is the incidence matrix between the $k$-dimensional sub-simplices $\sigma^k$ and $\sigma^{k+1}$ in a simplicial complex $T$,
\begin{equation}
    (I^{k+1}_k)_{ij} = 
        \begin{cases}
            \ \ 1& \textrm{if} \ \sigma^k_i \prec \sigma^{k+1}_j \quad \textrm{with same orientation},\\
            -1& \textrm{if} \ \sigma^k_i \prec \sigma^{k+1}_j \quad \textrm{with opposite orientation},\\
            \ \ 0& \textrm{if} \ \sigma^k_i \ \slashed{\prec} \ \sigma^{k+1}_j.
        \end{cases}
\end{equation}
The resulting operator $L^{(k)}$ is real and semi-positive because of its structure in terms of $I^{k+1}_k$ and its transpose. The eigenvectors corresponding to non-zero eigenvalues of $L^{(k)}$ are the union of the eigenvectors corresponding to non-zero eigenvalues of $L^{(k)}_+$ and $L^{(k)}_-$, which are orthogonal. The results that are relevant in the remainder of this text are invariant under changes of the orientation. For definiteness, we will, however, give an explicit prescription for choosing an orientation for all simplices in $T$. We label all vertices in $T$ with a unique integer. All $k$-simplices $\sigma^k$ can then be represented by their $k+1$ vertices $\sigma_i^0$, i.e. $\sigma^k=[\sigma^0_1, \ldots, \sigma^0_{k+1}]$. The orientation of $\sigma^k$ is determined by the order of its vertices. We choose to order them in increasing order of their labels, i.e. $\sigma^0_1 < \ldots < \sigma^0_{k+1}$. A $(k+1)$-simplex $\sigma^{k+1}=[\sigma^0_1, \ldots,\sigma^0_{k+2}]$ that contains the $k$-simplex $\sigma^k=[\sigma^0_1, \ldots,\slashed{\sigma}^0_i, \ldots,\sigma^0_{k+2}]$ are of the same orientation if $(-1)^{(i-1)}=1$ and of opposite orientation if $(-1)^{(i-1)} = -1$. The vertex $\slashed{\sigma}^0_i$ denotes the vertex that is part of $\sigma^{k+1}$, but not part of $\sigma^k$.

Besides making a choice for a discrete analogue of $\Delta^{(k)}$, we will have to make a choice how to treat the diffusion time $\tau$ to define a discrete analogue of the heat equation in Eq. \eqref{kHE}. For numerical reasons we will choose a discrete time variable $\tau$ and write $\tau = st$, where $s$ is a positive real number and $t$ is an integer time. The discrete heat equation will be defined on a $\tau$-dependent discrete $k$-form $v(\tau) \in \Omega^k(T)$. The state $v(\tau)$ at time $\tau$ can be represented by a vector of dimension $N_k$. Replacing the time derivative in Eq. \eqref{kHE} by a finite difference equation in $t$ and replacing $\Delta^{(k)}$ by $L^{(k)}$, the discrete heat equation is given by a finite difference equation in terms of the vector $v(\tau)$,
\begin{equation}
v(\tau + s)=(\mathbb{I}-sL^{(k)})v(\tau),
\end{equation}
where $\mathbb{I}$ is the unit matrix of dimension $N_k$. The equation for $v(\tau)$ can be solved iteratively. The iterations will be numerically stable if $s^{-1}\leq2\lambda_{max}$, where $\lambda_{max}$ is the largest eigenvalue of $L^{(k)}$. Iteratively acting with $A\equiv\mathbb{I}-sL^{(k)}$, the state $v(\tau)$ is given by,
\begin{equation}
v(\tau)=A^{t} v(0), \quad \tau = s t.
\label{StateDiff}
\end{equation}
The vector $v(0)$ at $\tau=0$ can be decomposed in the orthonormal eigenbasis $v_m$ of $L^{(k)}$,
\begin{equation}
v(0)=\sum_{m}c_m v_m, \quad L^{(k)}v_m=\lambda_mv_m.
\end{equation}
We have chosen\footnote{In previous studies of spectral dimensions in CDT \cite{SpecDimCDT2}, $s^{-1}$ was chosen to be equal to the number of neighbours of the simplices. These studies where conducted for the case $k=n$, so the number of neighbours was the same for every simplex. The discretisation artefact in \cite{SpecDimCDT2} of the running spectral dimensions for even and odd diffusion times for small $t$ is no longer present with the choice $s^{-1}=\lambda_{max}$.} $s^{-1}=\lambda_{max}$, such that the eigenvalues of $A$ lie between $0$ and $1$. The eigenvalues of $A$ equal to $1$ correspond to the zero modes of $L^{(k)}$. The diffusion will therefore converge to the subspace of zero modes of $L^{(k)}$ for large $\tau$,
\begin{equation}
\lim_{\tau\to\infty} v(\tau) = \sum_{m=1}^{m_0} (v(0)\cdot v_m)v_m,
\label{state}
\end{equation}
where $m_0$ is the number of zero modes and $v_m,\ m \in \{1,...,m_0\}$, is the orthonormal set of eigenvectors that spans the subspace of zero modes of $L^{(k)}$. The inner product $v(\tau) \cdot v_m$ is a conserved quantity for $m \in \{1,...,m_0\}$.

We can now define the discrete return probability $P_i^{(k)}(\tau)$ for a simplex $\sigma^k_i$. The discrete return probability is obtained by taking the orthonormal position basis vectors $e_i$ as the initial state, i.e. $v(0)=e_i$ and projecting $v(\tau)=A^{t}v(0)$ on the same initial state $e_i$,
\begin{equation}
P_i^{(k)}(\tau) = (A^{t} e_i) \cdot e_i \, .
\label{RetProb}
\end{equation}
The basis vectors $e_i$ are the analogue of the delta function distributions of equation \eqref{delta} and are associated with the $k$-simplex $\sigma^k_i$. The average return probability $P^{(k)}$ is then given by
\begin{equation}
P^{(k)}(\tau) = \frac{1}{N_k}\sum_i P_i^{(k)}(\tau) = \frac{1}{N_k}\mathrm{Tr} \ A^{t}, \quad \tau=s t.
\end{equation}
The trace of the operator $A^t$ can be expressed in terms of its eigenvalues, so we conclude that,
\begin{equation} \label{eq:d3.22}
P^{(k)}(\tau) = \frac{1}{N_k}\sum_{m,n} (1-s\lambda^m_n)^{\tau/s}.
\end{equation}
We now have an explicit $\tau$-dependent expression for the discrete spectral dimension $D^{(k)}$ for $k$-forms,
\begin{equation}
D^{(k)}(\tau)= -2 \tau \frac{\Delta \log P^{(k)}(\tau)}{\Delta \tau} =- 2(\tau/s) \log(P^{(k)}(\tau+s)/P^{(k)}(\tau)),
\label{SpecDim}
\end{equation}
where $\Delta$, in this equation, is the finite difference operator for a discrete diffusion time step $\tau$. If we want that the discrete diffusion time step is approximately equal to a continuum diffusion time we require,
\begin{equation}
   | \sum_i (1-\lambda_i)^\tau - \sum_i e^{-\tau\lambda_i}| \ll 1, \ \mathrm{for}, \ \tau \gg \log N_k.
\end{equation}
We can therefore expect that the artefacts of discretising the diffusion time are small for $\tau \gg \log N_k$ and we expect that we can measure regularisation invariant properties of expectation values of the spectral dimension $\langle D^{(k)} \rangle$ for these timescales. The largest eigenvalues of $L^{(k)}$ are associated to length scales close to the lattice spacing and are considered to be regularisation artefacts in the context of the gravitational path-integral. For longer diffusion times $\tau$ the process will explore a larger portion of the triangulation $T$. We will interpret the spectral dimension $D^{(k)}$ at larger $\tau$ as the effective dimension at a length scale $\ell_\tau$ associated to the timescale $\tau$. In the continuum case, a normalisation of $\Delta^{(k)}$ corresponds to a rescaling of the diffusion time $\tau$ with respect to the volume of the geometry $V$, as was shown in equation \eqref{RetScale}. To compare the timescale of the diffusion on different geometries of the same volume in the discrete case, we therefore have to take the scale $s$ of the diffusion time $\tau=st$ into account. When comparing the spectral dimensions $D^{(k)}(\tau)$ for different geometries of the same volume, we will therefore normalise the diffusion time $\tau$ by the factor $s$.

It is warranted to associate an effective dimension to timescale $\tau$ if the average return probability $P^{(k)}(\tau)$ (approximately) scales with a constant power. This will be the case if we have a timescale of the spectral dimension $D^{k}(\tau)$ at which the derivative is (approximately) zero. We will call such a regime of $D^{k}(\tau)$ a plateau. At such a timescale, $D^{k}(\tau)$ is constant up to linear order in some small expansion parameter $\epsilon$, so we can write,
\begin{equation}
    D^{k}(\tau) = c+\mathcal{O}( \epsilon).
\end{equation}
For a plateau around the value $c$, $P^{(k)}(\tau)$ scales like,
\begin{equation}
    P^{(k)}(\tau) \sim \tau^{-c/2} + \mathcal{O}(\epsilon).
    \label{SpecDimForm}
\end{equation}
We therefore conclude that we can interpret the presence of a plateau in the spectral dimension at a timescale $\tau$ as signalling an effective dimension equal to $c$ in the triangulation $T$ as long as $D^{k}(\tau)$ changes slowly in comparison to $c$.

We can now compute the discrete spectral dimension $D^{(k)}$ by computing the averaged return probability with the eigenvalues of $L^{(k)}$, which is equivalent to computing the trace of $A^{t}$. For numerical studies of large geometries, it is necessary to approximate the averaged return probability $P^{(k)}(\tau)$. This can be done by sampling the return probability $P^{(k)}_i(\tau)$ given by equation \eqref{RetProb} for a subset of initial $k$-simplices $\sigma^k_i$. For the cases $k=0$ and $k=n$, the return probability $P^{(k)}_i(\tau)$ can be interpreted in terms of the probability that a random walk that starts at $\sigma^k_i$ returns to $\sigma^k_i$ after a time $\tau$. Equivalently, for $k=0$ and $k=n$, $P^{(k)}_i(\tau)$ can be interpreted in terms of the amplitude of a scalar  field at $\sigma^k_i$ that initially has a delta function distribution and is subjected to a diffusion process. For other values of $k$, we will interpret $P^{(k)}_i(\tau)$ in analogy to the scalar cases.

\begin{figure}[h!]
\centering
\includegraphics[width=0.56\textwidth]{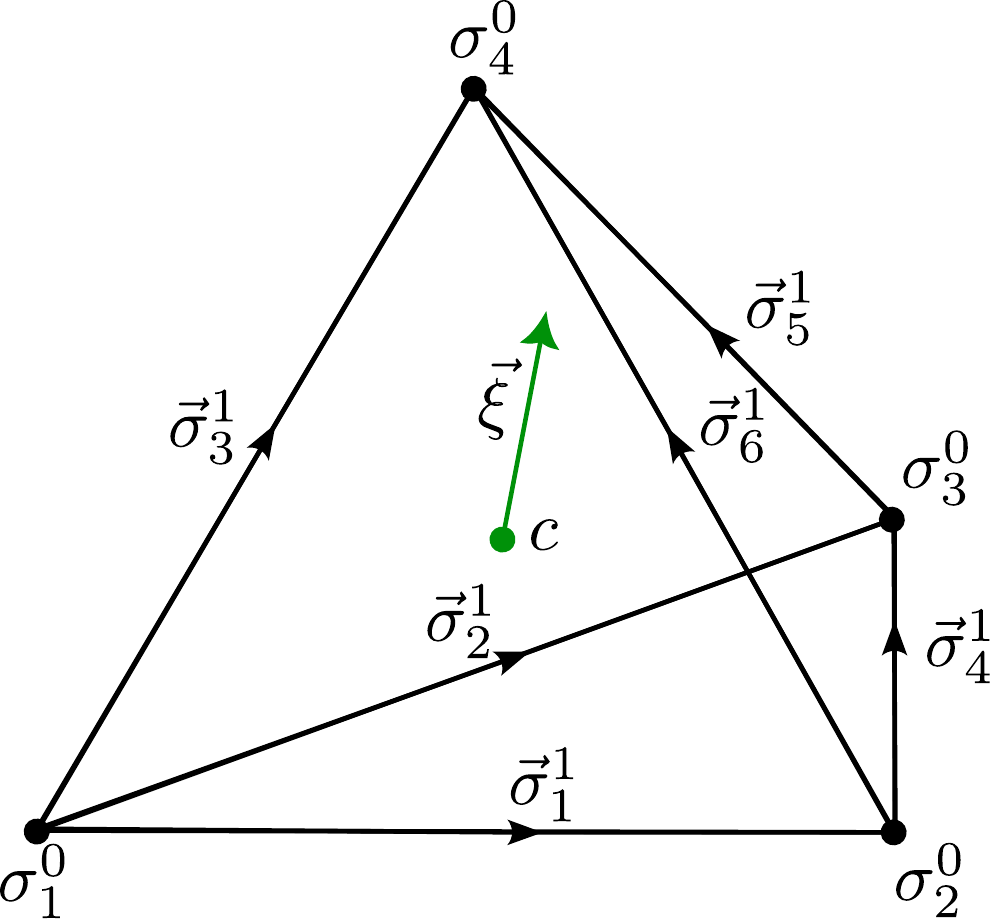}
\caption{The figure shows the four vertices $\sigma_i^0$ of a tetrahedron $\sigma^3$ with its link vectors $\vec{\sigma}_i^1$. Given a one-form field $\alpha^1$, the corresponding vector field evaluated on $\sigma^3$ is given by $\vec{\xi}=\alpha^\sharp(\sigma^3)=\sum^6_{i=1} \alpha(\sigma^1_i) \vec{\sigma}^1_i$.}
\label{VFD2}
\end{figure}

\begin{figure}[H]
\begin{minipage}[h]{0.47\linewidth}
\begin{center}
\includegraphics[width=.2\linewidth]{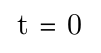}
\end{center}
\end{minipage}
\hfill
\begin{minipage}[h]{0.47\linewidth}
\begin{center}
\includegraphics[width=.2\linewidth]{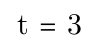}
\end{center}
\end{minipage}
\vspace*{0.5cm}
\vfill

\begin{minipage}[h]{0.47\linewidth}
\begin{center}
\includegraphics[width=.9\linewidth]{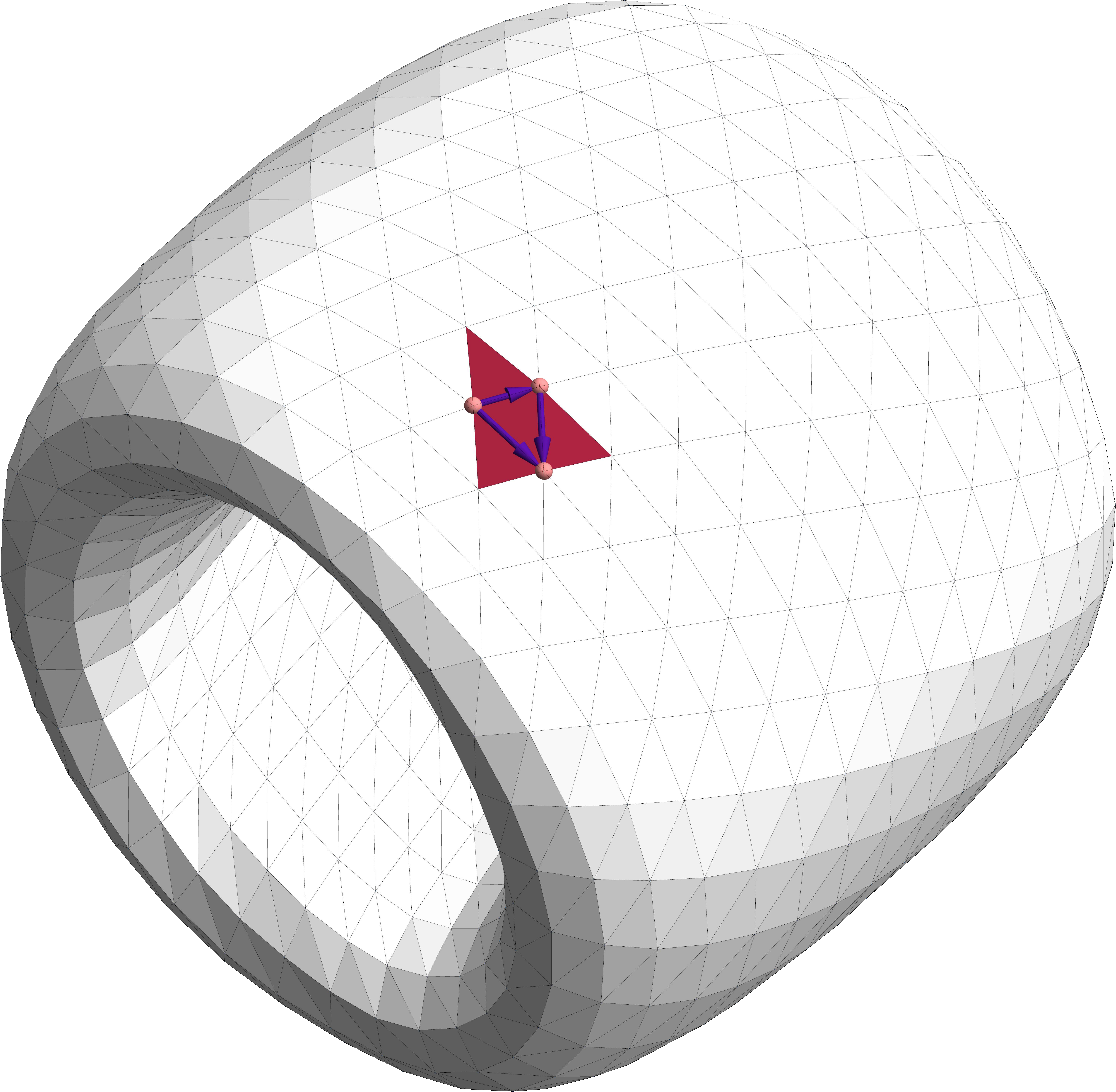}
\end{center}
\end{minipage}
\hfill
\begin{minipage}[h]{0.47\linewidth}
\begin{center}
\includegraphics[width=.9\linewidth]{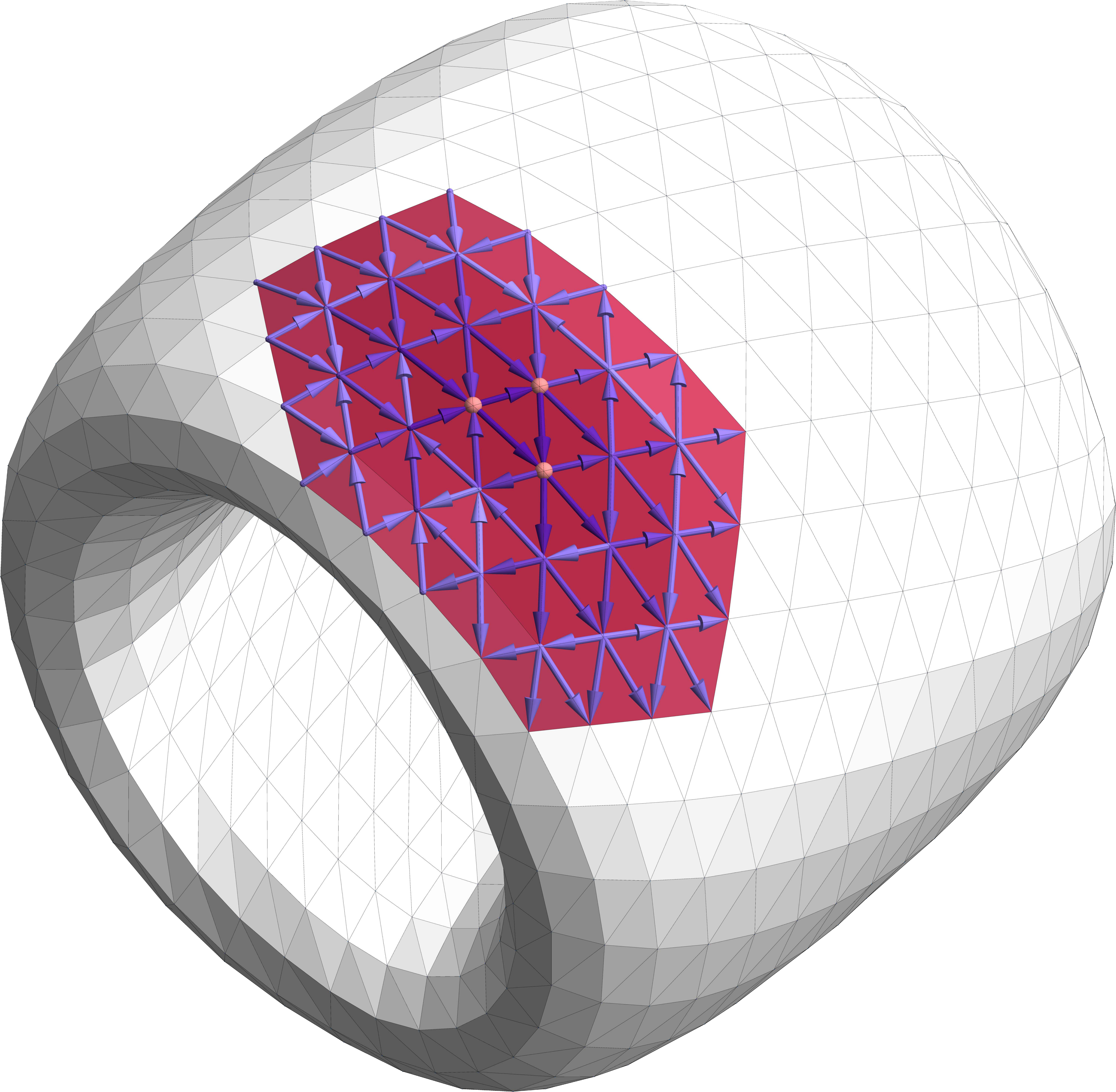}
\end{center}
\end{minipage}
\vspace*{0.5cm}
\vfill

\begin{minipage}[h]{0.47\linewidth}
\begin{center}
\includegraphics[width=.2\linewidth]{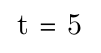}
\end{center}
\end{minipage}
\hfill
\begin{minipage}[h]{0.47\linewidth}
\begin{center}
\includegraphics[width=.2\linewidth]{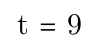}
\end{center}
\end{minipage}
\vspace*{0.5cm}
\vfill

\begin{minipage}[h]{0.47\linewidth}
\begin{center}
\includegraphics[width=.9\linewidth]{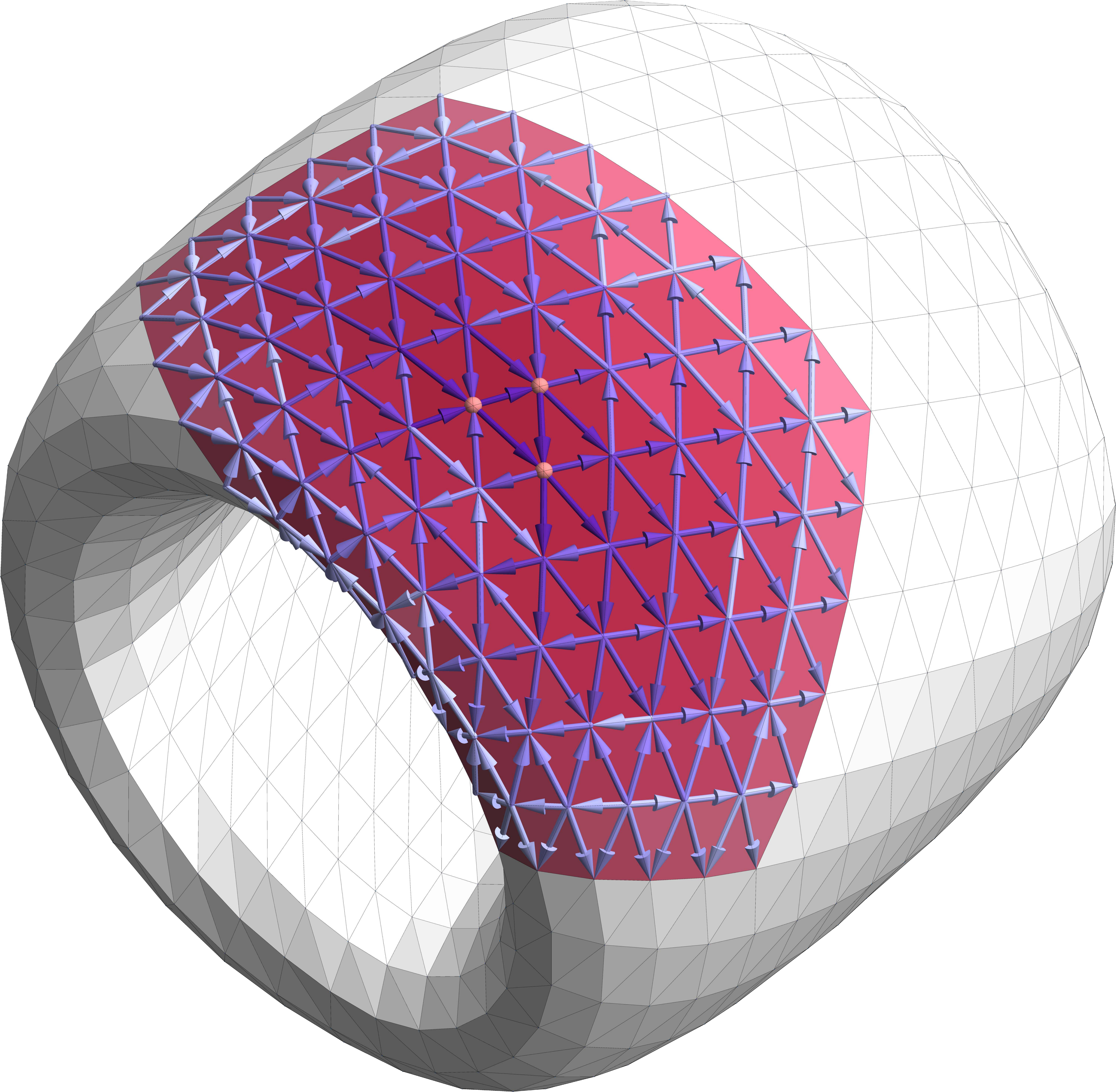}
\end{center}
\end{minipage}
\hfill
\begin{minipage}[h]{0.47\linewidth}
\begin{center}
\includegraphics[width=.9\linewidth]{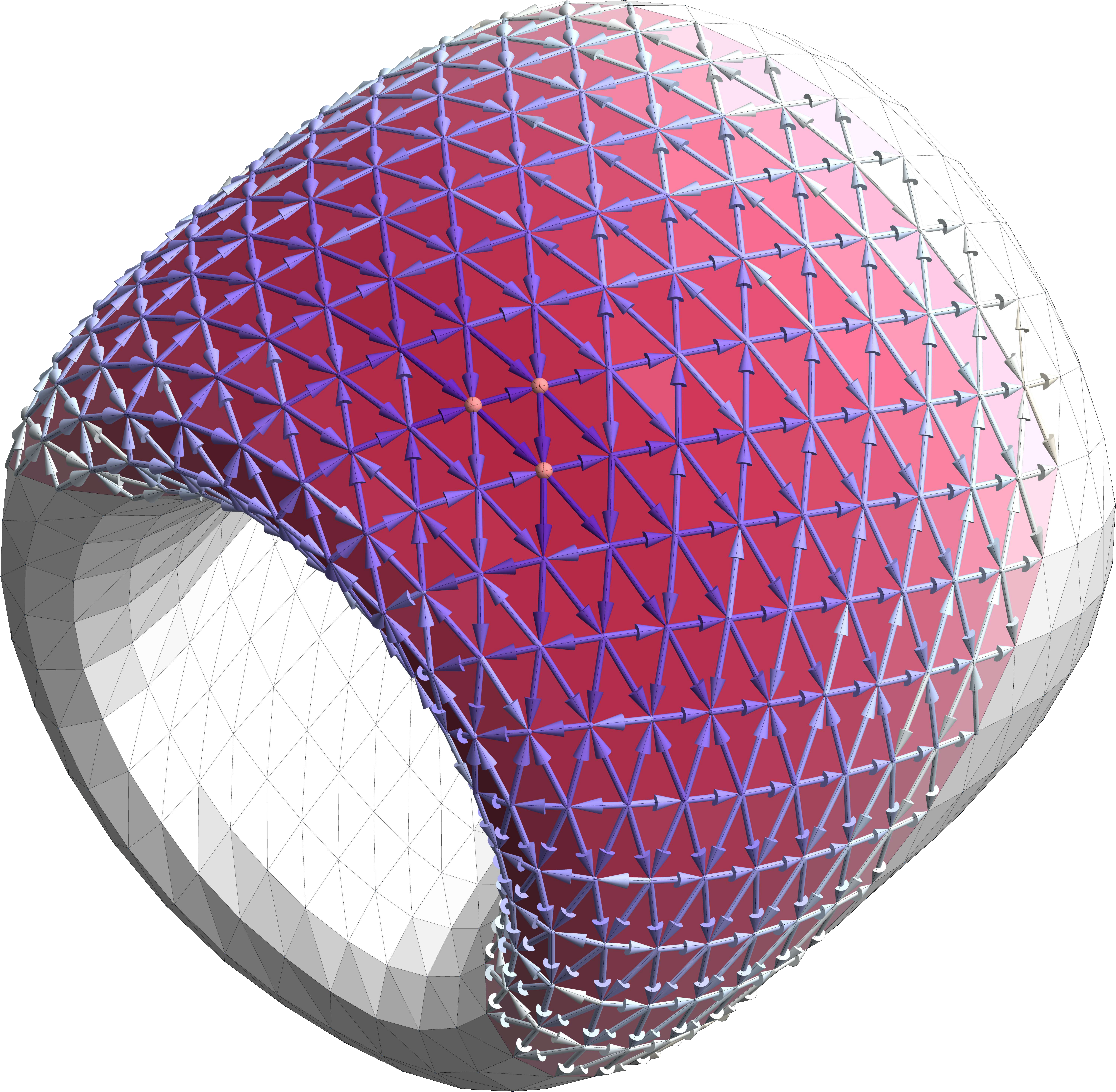}
\end{center}
\end{minipage}
\vspace*{0.5cm}
\vfill
\begin{minipage}[h]{1\linewidth}
\begin{center}
\includegraphics[width=0.8\linewidth]{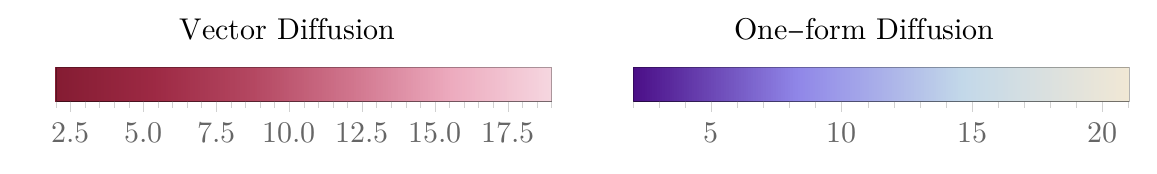}
\end{center}
\end{minipage}
\caption{The figure shows the diffusion process of a discrete one-form field $\alpha^1$ and a discrete vector field $X^1$, both governed by the one-form heat equation on a flat two-torus consisting of $30\times 30$ vertices. The links are coloured according to $-1$ times the log of norm of the one-form field $\alpha^1(\sigma^1_i)$ evaluated on the link $\sigma^1_i$ and the triangles are coloured according to $-1$ times the log of the norm of the vector field $||\xi(\sigma^2_i)||$ evaluated on the triangle $\sigma^2_i$, both as a function of $\tau$. The four images refer to the diffusion steps $t=0$ (top-left), $t=3$ (top-right), $t=5$ (bottom-left), $t=9$ (bottom-right), $\tau=st$. The initial state of $X^1$ is concentrated on the triangle indicated by the three orange dots. The initial state of $\alpha^1$ is concentrated on one of the links of the triangle indicated by the three orange dots. At the locus of their respective initial states, the norm of $\alpha^1$ is equal to $P^1(\tau)$ and the norm of $X^1$ is equal to $\vec{P}(\tau)$.}
\label{diffusion}
\end{figure}

For a differentiable manifold, $k$-form fields and $k$-tensor fields are dual to each other. As is often the case for generalisations of continuum objects to discrete manifolds, there is no unique choice for the definition of discrete $k$-form fields and discrete $k$-tensor fields. A specific definition for both objects has been studied in detail in \cite{MCA1}. The $k$-form fields were already discussed above and are associated to the $k$-simplices. The $k$-tensor fields $X^k \in \mathfrak{X}^k(T)$ however, are elements of $\otimes^k \mathbb{R}^n$. Noting that the tangent space is not well defined on the boundary of a $n$-simplex, it is unnatural to define $X^k$ on the $k$-simplex $\sigma^k$ for the cases $0<k<n$. Instead, it will be defined on the interior of the $n$-simplices, where there exists a unique definition of a tangent space. In \cite{MCA1} it was discussed how defining discrete vector fields on the interior of simplices has certain advantageous properties. It is important to note that discrete $k$-tensor fields defined in this fashion do not admit a duality relation to the discrete $k$-form fields, because the number of $k$-simplices for $0<k<n$ is different from the number of $n$-simplices for simplicial manifolds $T$. However, there exists a surjective function\footnote{This definition is similar to what is known as a Whitney form \cite{bossavit1998}, \cite{hirani_discrete_2003}, which has been shown to have good convergence properties to continuum tensor fields.} $\sharp: \Omega^k(T)\rightarrow \mathfrak{X}^k(T)$, 
\begin{equation}
(\alpha^k)^\sharp(\sigma^n_i) = \sum_{\sigma^k_j \prec \sigma_i^n } \alpha^k(\sigma^k_j) \vec{\sigma}_j^k,
\end{equation}
where $(\alpha^k)^\sharp(\sigma^n_i)$ is the value (in $\otimes^k \mathbb{R}^n$) of the discrete vector field $X^k=(\alpha^k)^\sharp$ in the simplex $\sigma^n_i$, the sum runs over all simplices $\sigma^k_j$ contained in the simplex $\sigma_i^n$, $\alpha^k(\sigma^k_j)$ is the value (in $\mathbb{R}$) of the discrete $k$-form evaluated on the simplex $\sigma^k_j$ and $\vec{\sigma}_j^k$ is the $k$-tensor associated to the simplex $\sigma^k_j$. We can also define the injective function $\flat: \mathfrak{X}^k(T) \rightarrow \Omega^k(T)$,
\begin{equation}
(X^k)^\flat(\sigma^k_j) = \sum_{\sigma^n_i\succ \sigma_j^k} X^k(\sigma_i^n)\cdot \vec{\sigma}^k_j,
\end{equation}
where $(X^k)^\flat(\sigma^k_j)$ is the value (in $\mathbb{R}$) of the $k$-form $\alpha^k=(X^k)^\flat$ evaluated on $\sigma^k_j$, the sum runs over all simplices $\sigma^n_i$ which share the simplex $\sigma_j^k$, $X^k(\sigma^n_i)$ is the value (in $\otimes^k \mathbb{R}^n$) of the discrete vector field $X^k$ in the simplex $\sigma^n_i$ and $\cdot$ denotes the standard inner product defined on the tangent space of the flat interior of $\sigma_i^n$. The two functions $\sharp$ and $\flat$ connect the two spaces $\Omega^k(T)$ and $\mathfrak{X}^k(T)$. The sums in these equations run over the $k$-simplices $\sigma^k_j$ contained in the $n$-simplices $\sigma^n_i$ and the $n$-simplices $\sigma^n_i$ that contain the $k$-simplices $\sigma^k_j$, respectively. The vector $\vec{\sigma}^k_j$ and the dot product are defined with respect to the tangent space of the flat interior of $\sigma^n_i$. For the case $k=1$, $\mathfrak{X}^k(T)$ defines what is known as discrete vector fields, for which an example is given in figure \ref{VFD2}. Various studies show that discrete vector fields possess more favourable properties than discrete one-form fields when the goal is to approximate continuum vector fields or one-form fields \cite{Desbrun,Ben-Chen,MCA1}. In view of these deliberations, the question arises if a discrete notion of diffusion can be defined both in terms of discrete $k$-form fields and discrete $k$-tensor fields. The case of discrete $k$-form fields is given by equation \eqref{RetProb}. For diffusion of discrete $k$-tensor fields, we propose the following definition of vector field return probability $\vec{P}_i^{(k)}$. We consider a discrete unit-length vector field $X_i(0)$ which is zero everywhere except in one $n$-simplex $\sigma^n_i$. The vector field is then mapped to a discrete $k$-form field $f_i = (X_i)^\flat$ with the flat operator $\flat$. The resulting $k$-form field $f_i$ is multiplied $\tau$ times by the generator of diffusion $A$ and mapped back to the space of discrete vector fields with the sharp operator $\sharp$. The resulting discrete vector field $(A^t f_i)^\sharp$ is then projected on the initial $n$-simplex $\sigma^n_i$, which gives a single vector $\vec{\xi}(\tau)$ for which we compute the norm $||\vec{\xi}(\tau)||$. This defines the $k$-tensor return probability $\vec{P}_i^{(k)}$,
\begin{equation}
\vec{P}_i^{(k)}(\tau) = ||\vec{\xi}(\tau)||, \quad \vec{\xi}(\tau) = (A^t f_i)^\sharp(\sigma^n_i), \quad f_i = (X_i)^\flat, \quad \tau=st.
\end{equation}
where the index $i$ refers to the $n$-dimensional simplex $\sigma^n_i$. With the $k$-tensor return probability $\vec{P}_i^{(k)}(\tau)$ we define the $k$-tensor spectral dimension $\vec{D}^{(k)}$,
\begin{equation}
\vec{D}^{(k)}(\tau)=- 2\tau \log(\vec{P}^{(k)}(\tau+s)/\vec{P}^{(k)}(\tau)), \quad \vec{P}^{(k)}(\tau)=\frac{1}{N_n}\sum^{N_n}_i\vec{P_i}^{(k)}(\tau).
\label{SpecDimVec}
\end{equation}
In Fig. \ref{diffusion} we illustrate both the one-form diffusion $P^{(1)}(\tau)$ and vector diffusion $\vec{P}^{(1)}(\tau)$ for the relatively simple case of the flat two-torus.
\subsection{Two- and three-dimensional regular geometries}
\label{regular}
To interpret our results on generalised spectral dimensions in spatial slices of four-dimensional CDT, it is worthwhile to first discuss generalised spectral dimensions in a simplified setting. It is a difficult task to disentangle physically relevant information from discretisation effects appearing at small diffusion times and finite-size effects, which appear at large diffusion times. Both discretisation effects and finite size effects are considered regularisation artefacts. As an aid to interpret the measurements, it is useful to first investigate these effects on regular discrete geometries.

\begin{figure}[t!] 
    \centering
    \includegraphics[width=0.49\linewidth]{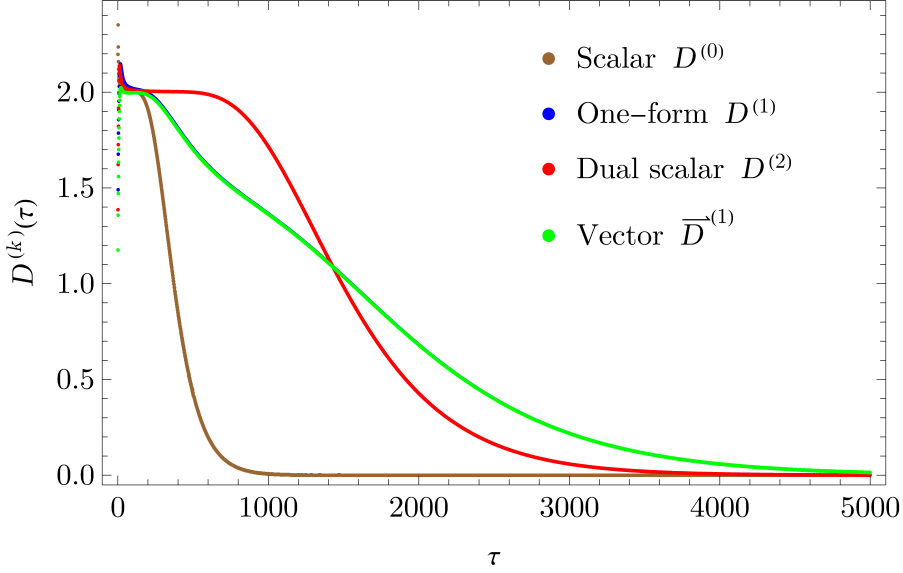}
    \includegraphics[width=0.49\linewidth]{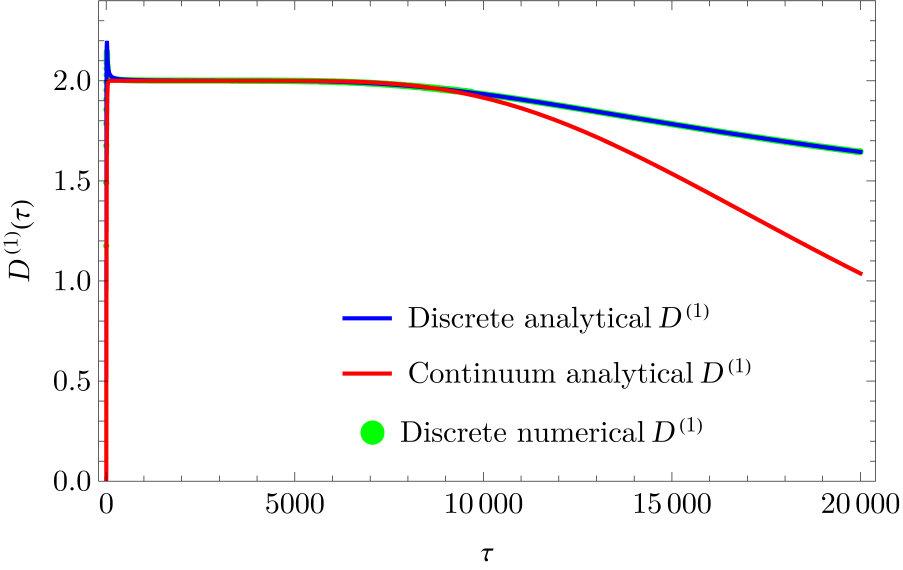}
    \caption{The figure to the left shows the generalised spectral dimensions for a triangulation of the flat two-dimensional torus consisting of $30 \times 30$ vertices. The spectral dimensions $D^{(0)}$, $D^{(1)}$ and $D^{(2)}$ differ significantly from 2 for small diffusion time $\tau$, due to discretisation effects. These effects are less pronounced for the vector spectral dimension $\vec{D}^{(1)}$, which has a value closer to $2$ for small $\tau$. An extended plateau appears for each of the spectral dimensions at a value $2\pm0.002$. The figure to the right shows $D^{(1)}$ from the analytical calculation for the flat torus presented in Appendix \ref{app} and the continuum one-form spectral dimension as a point of reference. It can be seen that the initial peak for small diffusion times is an artefact of the discretisation and the plateau is the feature that shows the dimension of the geometry.}
    \label{dfig:3proof-of-concept}
\end{figure}

We will consider a two-dimensional triangulation of the flat torus, which can be triangulated exactly. We will use Eqs. \eqref{SpecDim} and \eqref{SpecDimVec} to calculate the spectral dimensions for $k \in \{0,1,2\}$. Fig. \ref{dfig:3proof-of-concept} shows the four different spectral dimensions, the scalar spectral dimension $D^{(0)}(\tau)$, the one-form spectral dimension $D^{(1)}(\tau)$, the two-form spectral dimension $D^{(2)}(\tau)$ and the vector spectral dimension $\vec{D}^{(1)}(\tau)$, that appear for the two-dimensional flat torus. The figure to the left shows the spectral dimension curves obtained with numerical methods. The calculation  For $D^{(0)}(\tau)$, $D^{(1)}(\tau)$ and $D^{(2)}(\tau)$, we see significant deviations from the expected value of $2$ for small values of $\tau$. The deviation from $2$ is smallest for $\vec{D}^{(1)}(\tau)$. These results suggests that $\vec{D}^{(1)}(\tau)$ is less sensitive to discretisation effects.

An extended plateau at a value $2\pm0.002$ is present for each of the spectral dimensions. We therefore see that plateaus in the spectral dimension indeed indicate a range of time scales at which the effective dimension can be separated from discretisation artefacts and finite-size effects. The timescale at which the spectral dimension goes to zero varies for the different values of $k$. For different values of $k$, a discrete time step corresponds to a varying fraction of the total volume of the triangulation. It is therefore to be expected that the time scale at which finite-size effects become apparent depends on $k$. The curves in the figure to the right serve as a point of reference for the numerical results. These curves were obtained for $D^{(1)}(\tau)$ from an analytical calculation which is presented in Appendix \ref{app}. The continuum calculation of the one-form spectral dimension was obtained from the known spectrum of the continuum one-form Laplacian of the flat two-dimensional torus, which can be found in \cite{Craioveanu}. From the analytical calculations we can see that the peak for small diffusion time is indeed a discretisation artefact, because it is absent for the continuum case and that the plateau is the feature of the curve that is associated to the dimension of the geometry.

\begin{figure}[ht!]
\centering
\includegraphics[width=0.6\textwidth]{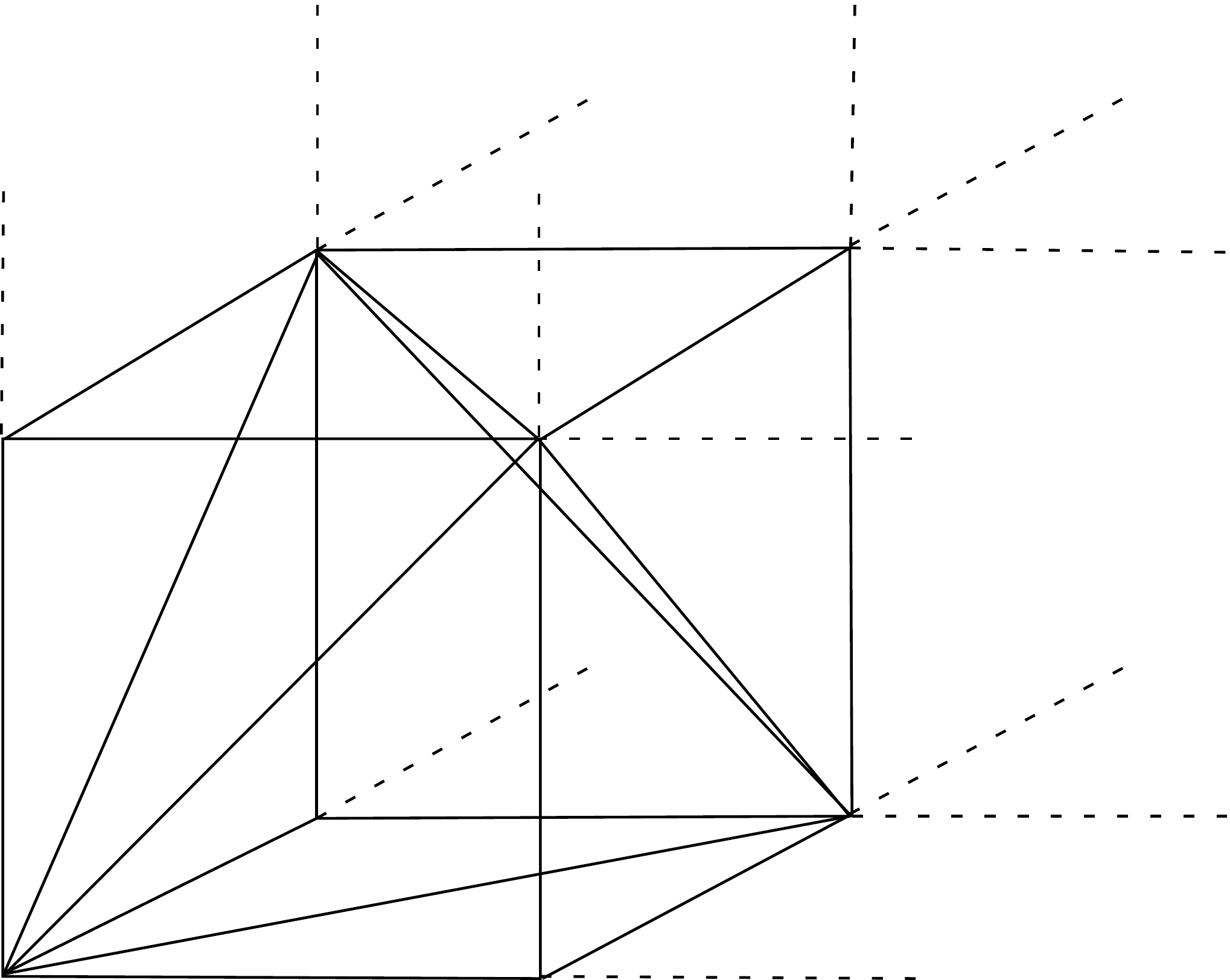}
\caption{An approximately flat equilateral three-dimensional torus can be constructed from repeatedly gluing copies (along the dashed lines) of the square block in this figure and identifying opposite sides. Every block consists of five equilateral tetrahedra. The blocks are glued in a way that links of different blocks coincide. All the links shown in the figure represent links of equal length. Such a block is not embeddable in a flat three-dimensional space.}
\label{SquareBlock}
\end{figure}

The CDT geometries we have investigated are equilateral, three-dimensional simplicial manifolds of toroidal topology. We will therefore discuss another regular geometry as a point of reference. An approximately flat three-dimensional torus can be constructed by gluing together square blocks of five tetrahedra such as the one illustrated in Fig. \ref{SquareBlock} and identifying opposite sides. The resulting geometry is only approximately flat, because flat three-dimensional space cannot be exactly triangulated by equilateral tetrahedra. The degrees of the vertices of the torus constructed by such square blocks are not homogeneous, as was the case for the flat two-dimensional torus. With this approximately flat torus we can investigate how such inhomogeneities impact the diffusion processes we are interested in.

\begin{figure}[ht!]
\includegraphics[width=0.49\textwidth]{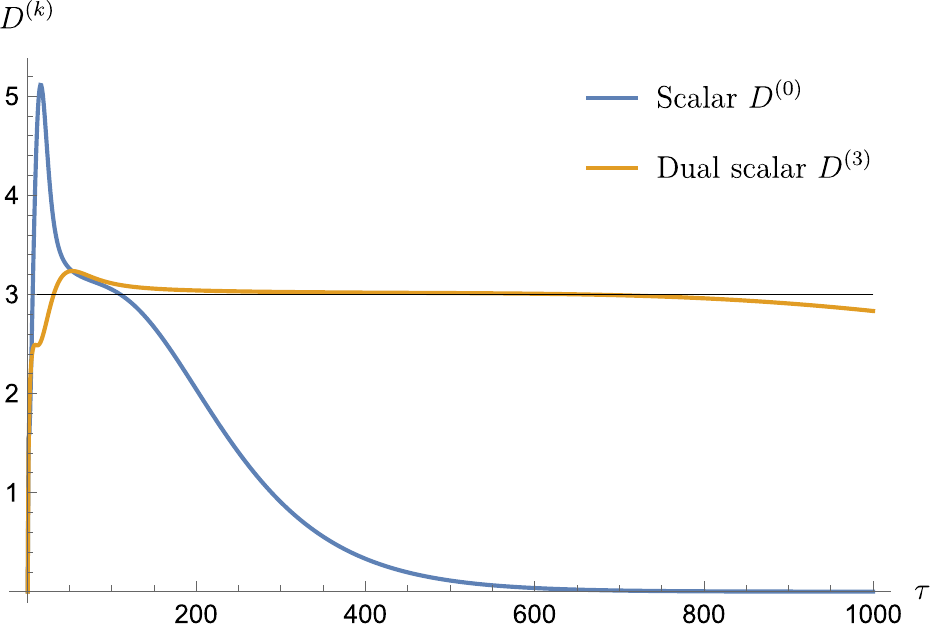}
\includegraphics[width=0.49\textwidth]{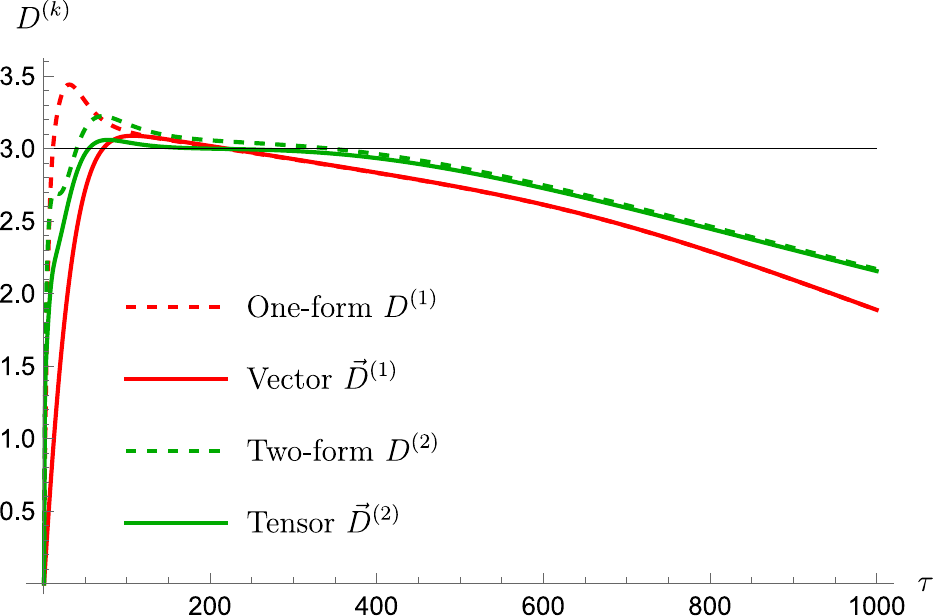}
\caption{The figures show the generalised spectral dimensions of an approximately flat three-dimensional torus. The approximately flat three-dimensional torus was constructed by gluing together blocks such as the one illustrated in Fig. \ref{SquareBlock} and identifying opposite sides. The torus consists of $4 \cdot 10^4$ tetrahedra.}
\label{FlatDims}
\end{figure}

On this approximately flat three-dimensional simplicial manifold we can calculate the k-form spectral dimensions $D^{(k)}$ with $k \in \{0,...,3\}$ and the tensor spectral dimensions $\vec{D}^{(k)}$ for $k \in \{1,2\}$. Fig. \ref{FlatDims} shows all the running spectral dimensions on the approximately flat torus for diffusion timescales before the zero modes become dominant (except for the scalar diffusion $D^{(0)}$). We note that all running spectral dimensions show, besides short time-scale effects, a plateau approximately equal to three as expected. Comparing the scalar spectral dimension $D^{(0)}$ to the dual scalar spectral dimension $D^{(3)}$, we see that there is a strong effect from the vertices of high degree, which cause the peak for small diffusion time. For larger diffusion times there is however an overlap between $D^{(0)}$ and $D^{(3)}$, showing that the inhomogeneities are washed out for larger diffusion times. There are significantly fewer vertices than tetrahedra (5 times fewer for the geometry of Fig. \ref{FlatDims}), so the finite-size effects are dominant much earlier for $D^{(0)}$, which diminishes the presence of the plateau.

Inspecting the curves for $D^{(1)}, D^{(2)}, \vec{D}^{(1)}$ and $\vec{D}^{(2)}$ we note that all curves show a plateau close to a value of three, but that the tensor spectral dimensions are less sensitive to small scale artefacts. The main conclusion we have drawn from investigating two and three-dimensional regular geometries, is that all spectral dimensions show a plateau at the expected value equal to the topological dimension for larger diffusion times. We have also shown that the various spectral dimensions are sensitive to inhomogeneities and lattice artefacts to a varying degree. The dual scalar spectral dimension $D^{(3)}$ and the $k$-tensor spectral dimensions $\vec{D}^{(k)}$ are particularly insensitive to lattice artefacts. Section \ref{Results} contains the measurements of the various spectral dimensions of spatial slices in the semi-classical phase $C$ of four-dimensional CDT.
\section{Generalised spectral dimensions in non-perturbative quantum gravity}
\label{Results} 
To investigate the various spectral dimensions in the context of non-perturbative quantum gravity we have generated CDT geometries with a Monte-Carlo algorithm based on the action given in equation \eqref{CDTAction}. We have chosen to study CDT geometries of toroidal topology, because in this case all Betti numbers are non-zero and all Laplace-Beltrami operators therefore have zero modes. The parameters we have chosen are $\kappa_0 = 4.0, \Delta = 0.2$ and the number of time slices is fixed to four. These parameters lie within the semi-classical phase $C$ of the CDT phase diagram (see Fig.~\ref{fig:my_label}) and have recently been studied in detail for toroidal topology \cite{topinduced}. We have also chosen this specific location in the CDT parameter space, because it is close to two triple points where various phase transition lines meet. As evidenced by the renormalization group studies \cite{Ambjorn:2014gsa,Renormalization_Ambjorn2020}, the triple points are potential candidates for a UV fixed point of CDT, however the existence of such a UV fixed point has not yet been definitively proven. It was shown that the effective lattice spacing decreases towards large $\kappa_0$ \cite{towardscontinum} (but depends slightly on $\Delta$). Low $\kappa_0$ seems to be the region of the parameter space related to the IR limit and high $\kappa_0$ (close to the triple points) corresponds to the UV limit of the theory. Our choice enables us therefore to investigate much finer effective lattice spacing and consequently quantum gravity effects related to much shorter physical distances than  the \emph{generic} point $(\kappa_0=2.2,\Delta=0.6)$, where most of CDT results in phase $C$ were so far obtained\footnote{There exists no explicit estimate of the effective lattice spacing for the case of CDT with toroidal spatial topology discussed here, but in the case of spherical topology the lattice spacing was estimated to change from around $2  \ \ell_{pl}$ (Planck lengths) for $\kappa_0=2.2$ to around $1 \ \ell_{pl}$ for $\kappa_0 = 4.4 $ \cite{towardscontinum}.}. At the same time we didn't want to get too close to the phase transition lines in order to be able to measure the IR properties of the quantum geometries. 

In order to extrapolate our results to the infinite volume limit, we have generated ensembles of geometries for the following target volumes: $\bar{N}_{41} = \{10\rk, 20\rk, 40\rk, 80\rk, 160\rk, 240\rk\}$. With these ensembles, we estimate the expectation value of the average return probabilities $\langle {P}_i^{(k)}(\tau) \rangle$ and $\langle \vec{P}_i^{(k)}(\tau) \rangle$,  as shown by equation \eqref{ExVal}, where we take ${P}_i^{(k)}(\tau)$ and $ \vec{P}_i^{(k)}(\tau)$ as the observable $\mathcal{O}(\mathcal{T})$.

For all generated four-dimensional geometries, we consider the three-dimensional spatial slices separately and calculate the average return probabilities ${P}_i^{(k)}(\tau)$ and $\vec{P}_i^{(k)}(\tau)$. Due to the finite computation time, the average return probability is calculated from a subset of the $k$-simplices in the three-dimensional slices. The average return probabilities are sampled by taking $2\times10^3$ randomly chosen $k$-simplices as the initial location of a diffusion process for each geometry. This number of initial locations gives a good approximation of the average return probability. Additional starting points give a contribution which is smaller than the quantum fluctuations present in the ensembles of geometries.

\begin{figure}[b!]
\centering
\includegraphics[width=0.49\textwidth]{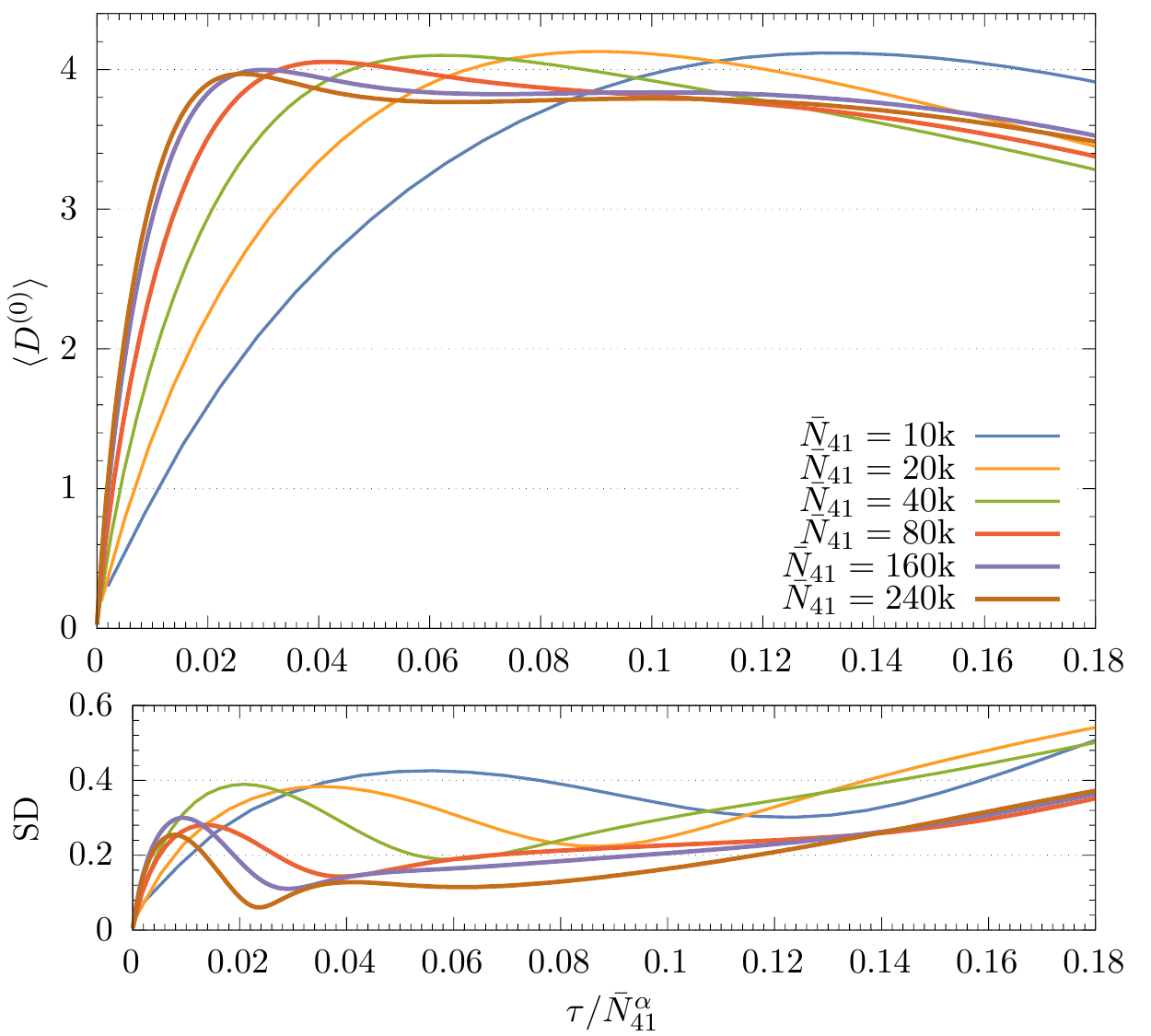}
\caption{The upper figure shows a sequence of curves with volume-normalised diffusion time $\tau/N^{\alpha}$ of the expectation value of the scalar spectral dimension $\langle D^{(0)} \rangle$ for geometries of different volumes. The volume of a CDT geometry is determined by the number of four-simplices, which is approximately equal to $\bar{N}_{41}$. The two curves for $\bar{N}_{41} = 160\rk$ and $240\rk$ are shown with a bold line. The parameter $\alpha=1.90$ was determined solely from the curves for which the plateaus are apparent. The lower figure shows the standard deviation $SD$ of the curves in the upper plot.}
\label{Points}
\end{figure}

\begin{figure}[H]
\centering
\includegraphics[width=0.49\textwidth]{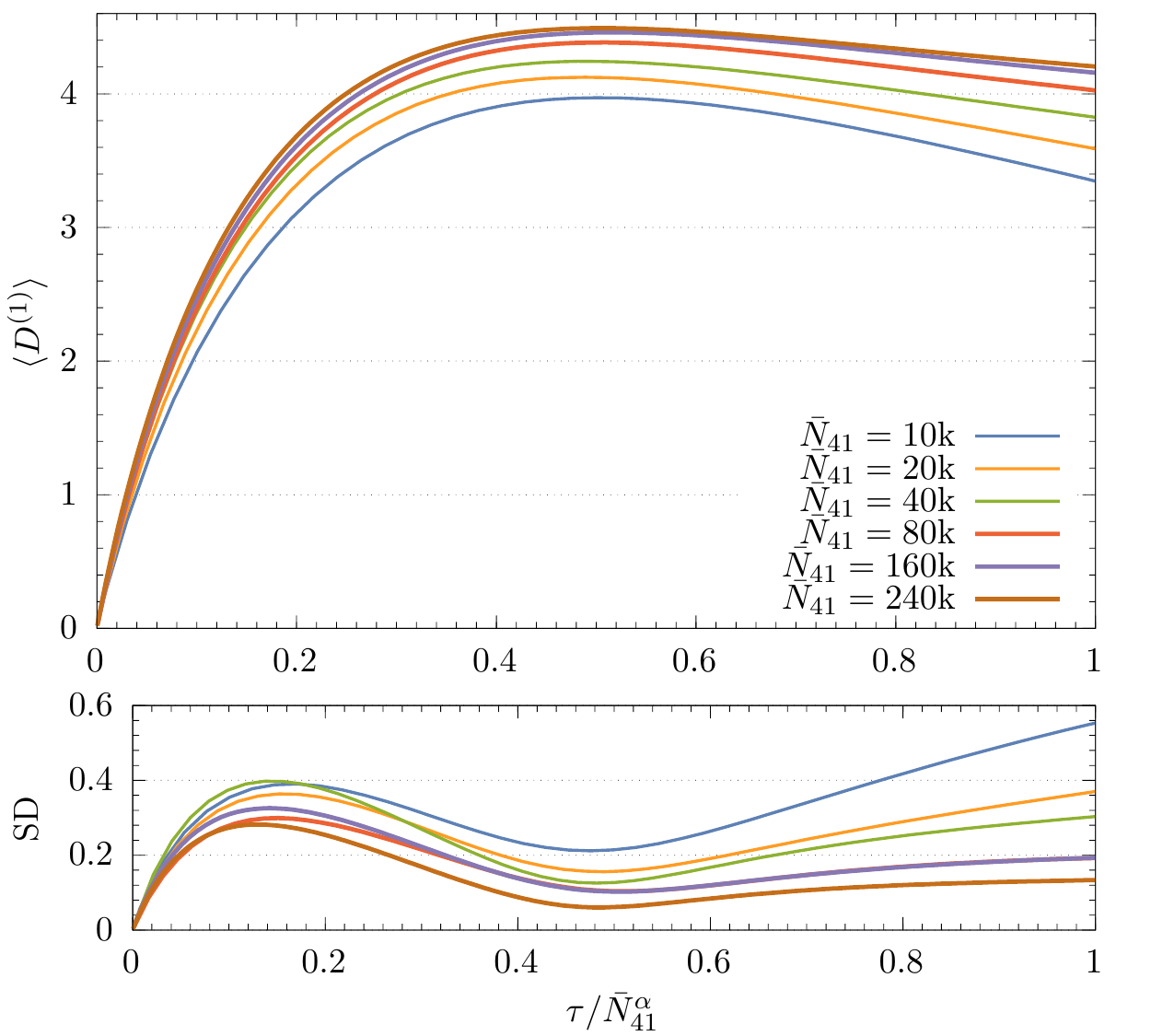}
\includegraphics[height=140 pt,width=0.49\textwidth]{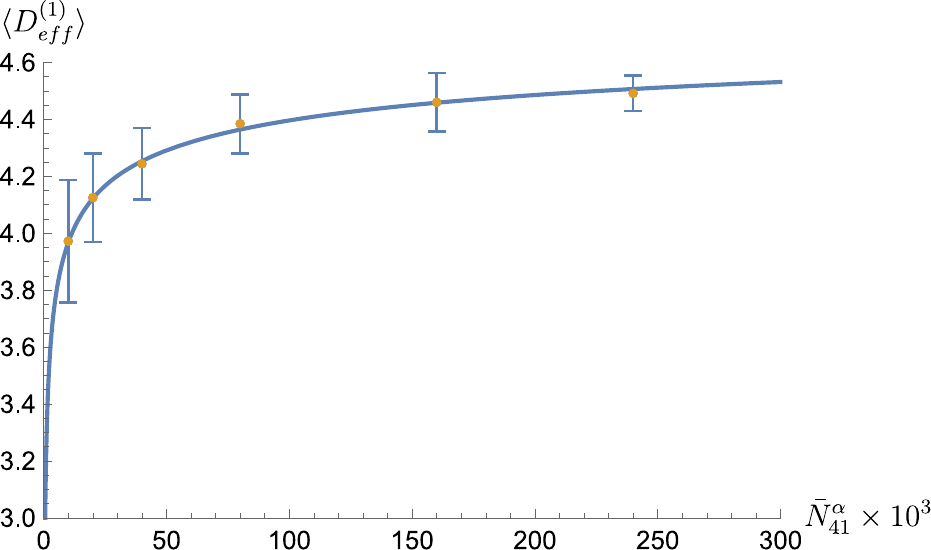}
\caption{The upper left figure shows a sequence of curves with volume-normalised diffusion time $\tau/N^{\alpha}$ of the expectation value of the one-form spectral dimension $\langle D^{(1)} \rangle$ for geometries of different volumes. The volume of a CDT geometry is determined by the number of four-simplices, which is approximately equal to $\bar{N}_{41}$. The two curves for $\bar{N}_{41} = 160\rk, 240\rk$ are shown with a bold line. The lower left figure shows the standard deviation $SD$ of the curves in the upper left plot. The right plot shows the finite-volume scaling of the maxima of the emerging plateaus of $\langle D^{(1)} \rangle$ as a function of $\bar{N}_{41}$. The same plot shows a power law fit of the form $a+b*\bar{N}_{41}^c$, from which we extrapolate an effective dimension $\langle {D_{eff}^{(1)}}
\rangle = 4.98 \pm 0.36$ in the infinite volume limit. The curves are shown for $\alpha=1.36$.}
\label{Links}
\centering
\includegraphics[width=0.49\textwidth]{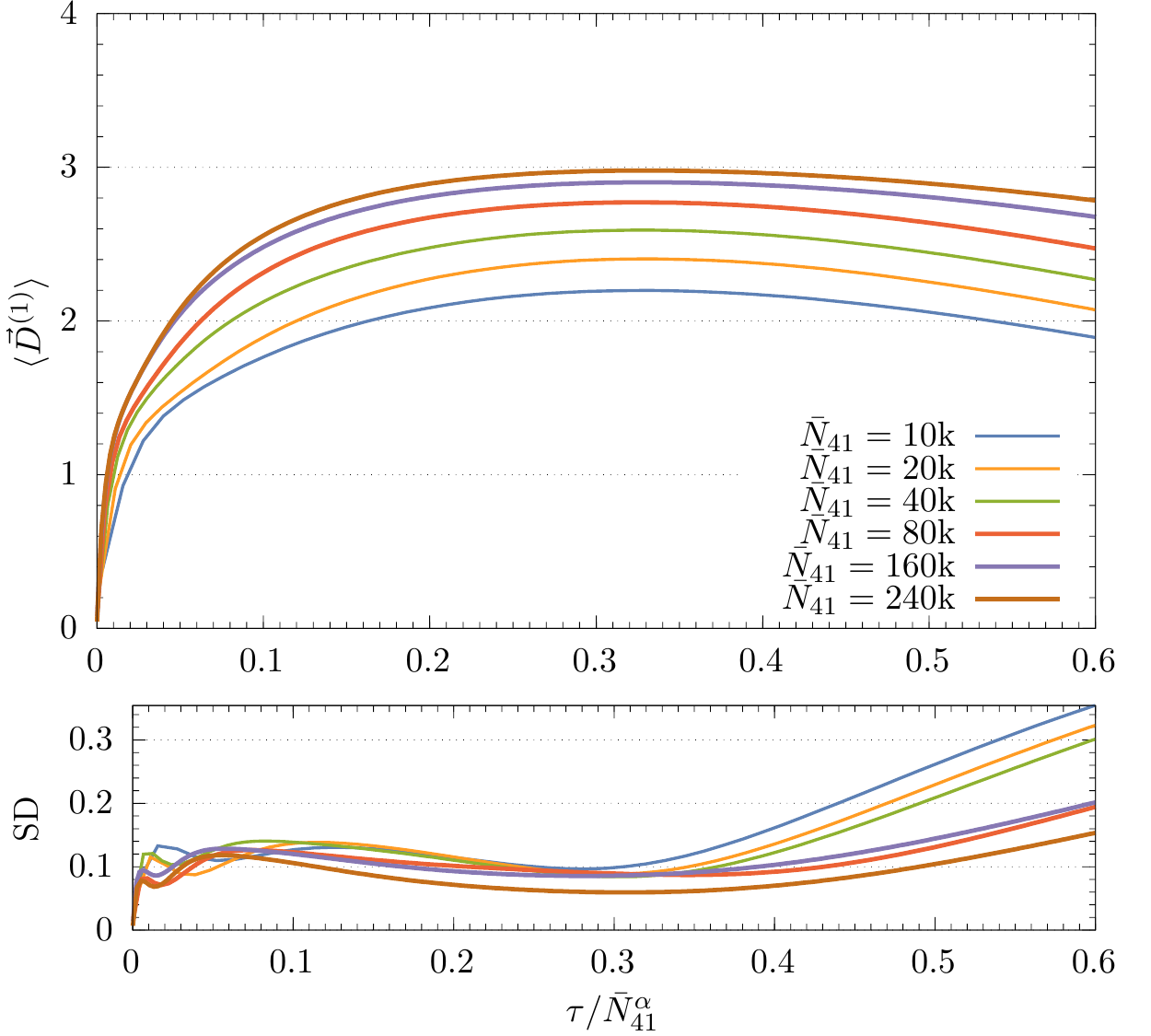}
\includegraphics[height=140 pt,width=0.49\textwidth]{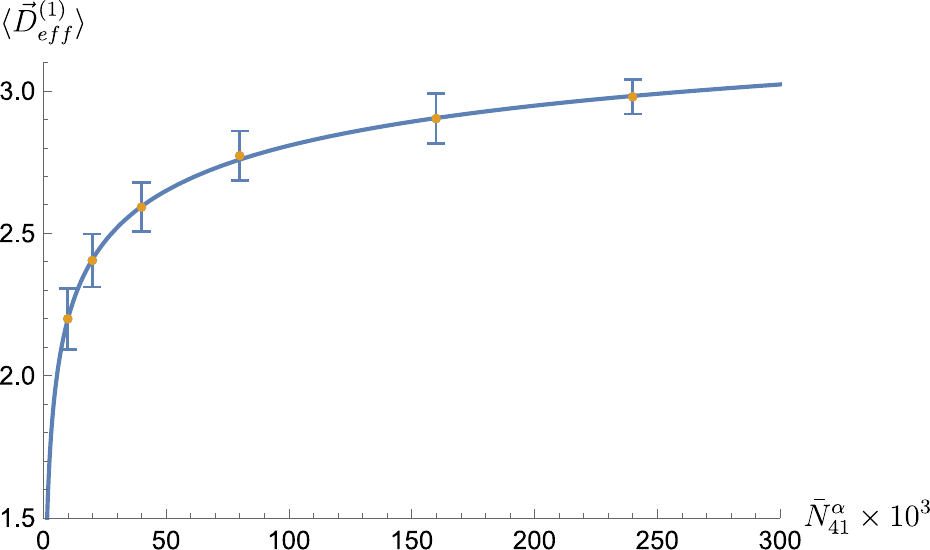}
\caption{The upper left figure shows a sequence of curves with volume-normalised diffusion time $\tau/N^{\alpha}$ of the expectation value of the vector spectral dimension $\langle \vec{D}^{(1)} \rangle$ for geometries of different volumes. The volume of a CDT geometry is determined by the number of four-simplices, which is approximately equal to $\bar{N}_{41}$. The two curves for $\bar{N}_{41} = 160\rk, 240\rk$ are shown with a bold line. The lower left figure shows the standard deviation $SD$ of the curves in the upper left plot. The right plot shows the finite volume scaling of the maxima of the emerging plateaus of $\langle \vec{D}^{(1)} \rangle$ as a function of $\bar{N}_{41}$. The same plot shows a power law fit of the form $a+b*\bar{N}_{41}^c$, from which we extrapolate an effective dimension $\langle {\vec{D}_{eff}^{(1)}}
\rangle = 4.03 \pm 0.45$ in the infinite volume limit. The curves are shown for $\alpha=1.65$.}
\label{LinksV}
\end{figure}

\begin{figure}[H]
\centering
\includegraphics[width=0.49\textwidth]{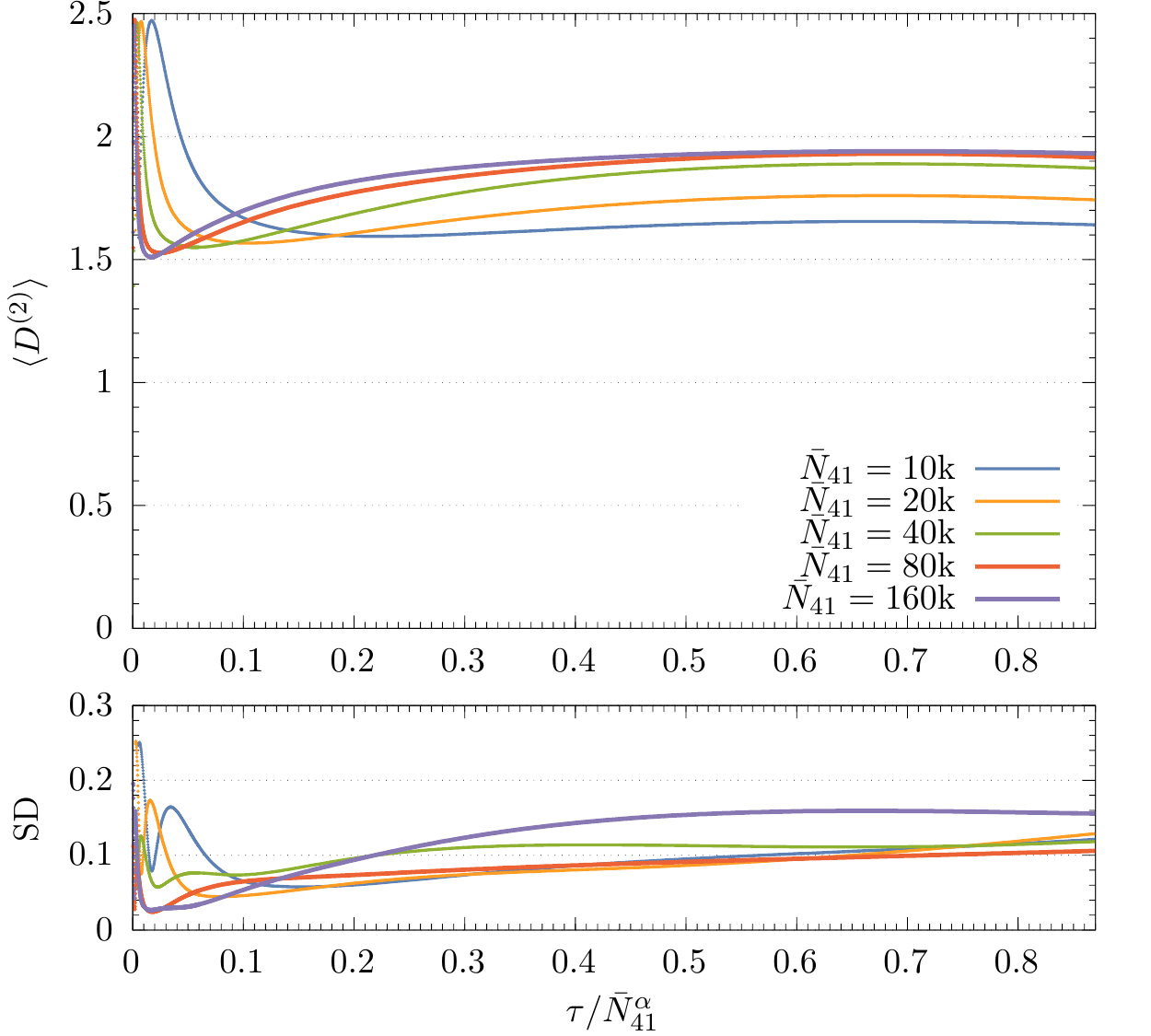}
\includegraphics[width=0.49\textwidth]{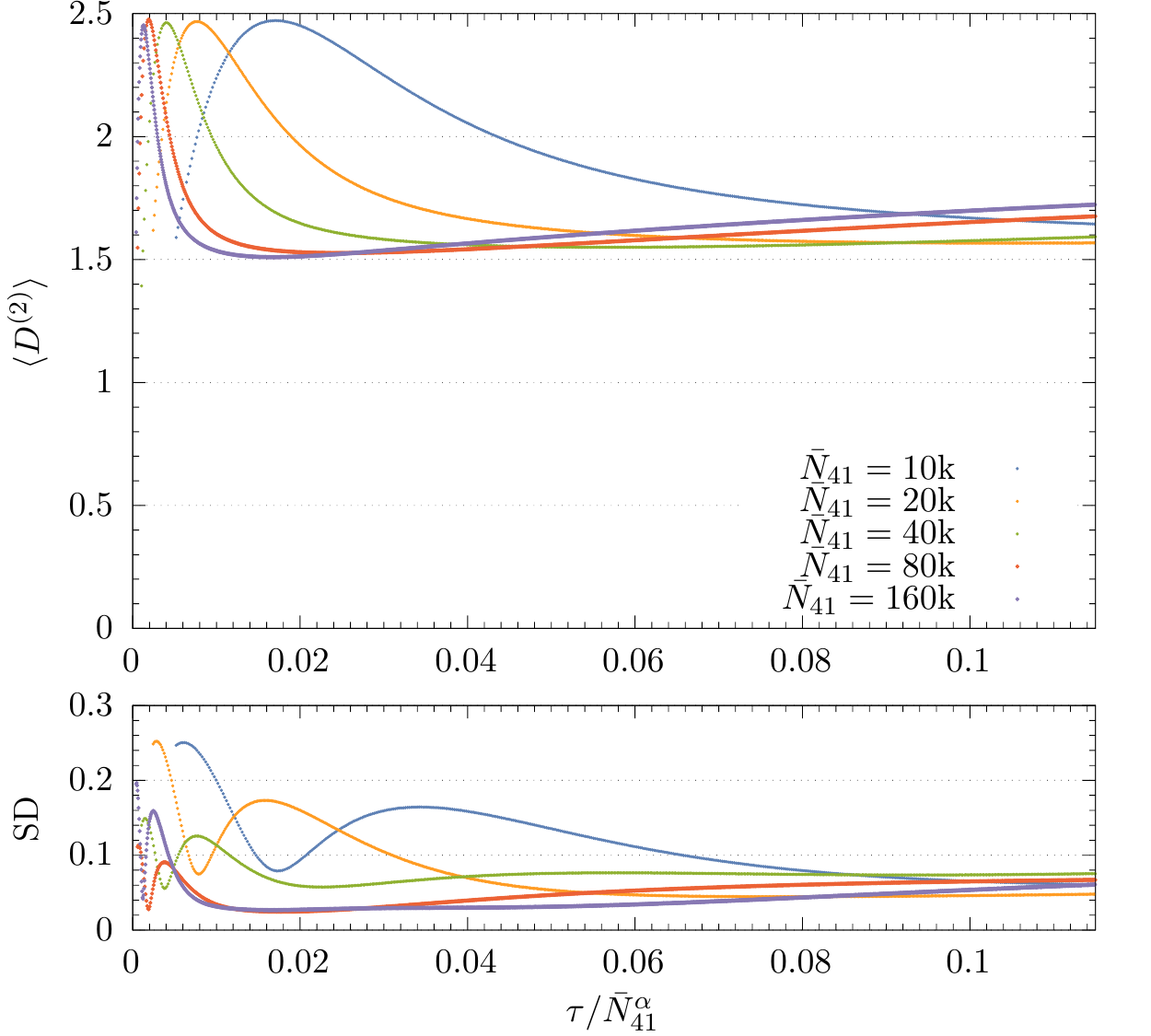}
\caption{The upper left figure shows a sequence of curves with volume-normalised diffusion time $\tau/N^{\alpha}$ of the expectation value of the two-form spectral dimension $\langle D^{(2)} \rangle$ for geometries of different volumes. The volume of a CDT geometry is determined by the number of four-simplices, which is approximately equal to $\bar{N}_{41}$. The two curves for $\bar{N}_{41} = 160\rk, 240\rk$ are shown with a bold line. The lower-left figure shows the standard deviation $SD$ of the curves in the upper-left plot. The upper right figure shows $\langle D^{(2)} \rangle$ for small diffusion times, and the lower right figure shows the corresponding standard deviation. The figures to the right show in detail how the lattice artefact peaks move to the left for increasing $\bar{N}_{41}$ and how the small diffusion time plateau emerges. The curves are shown for $\alpha=1.31$.}
\label{Triangle}
\end{figure}

\begin{figure}[H]
\centering
\includegraphics[width=0.49\textwidth]{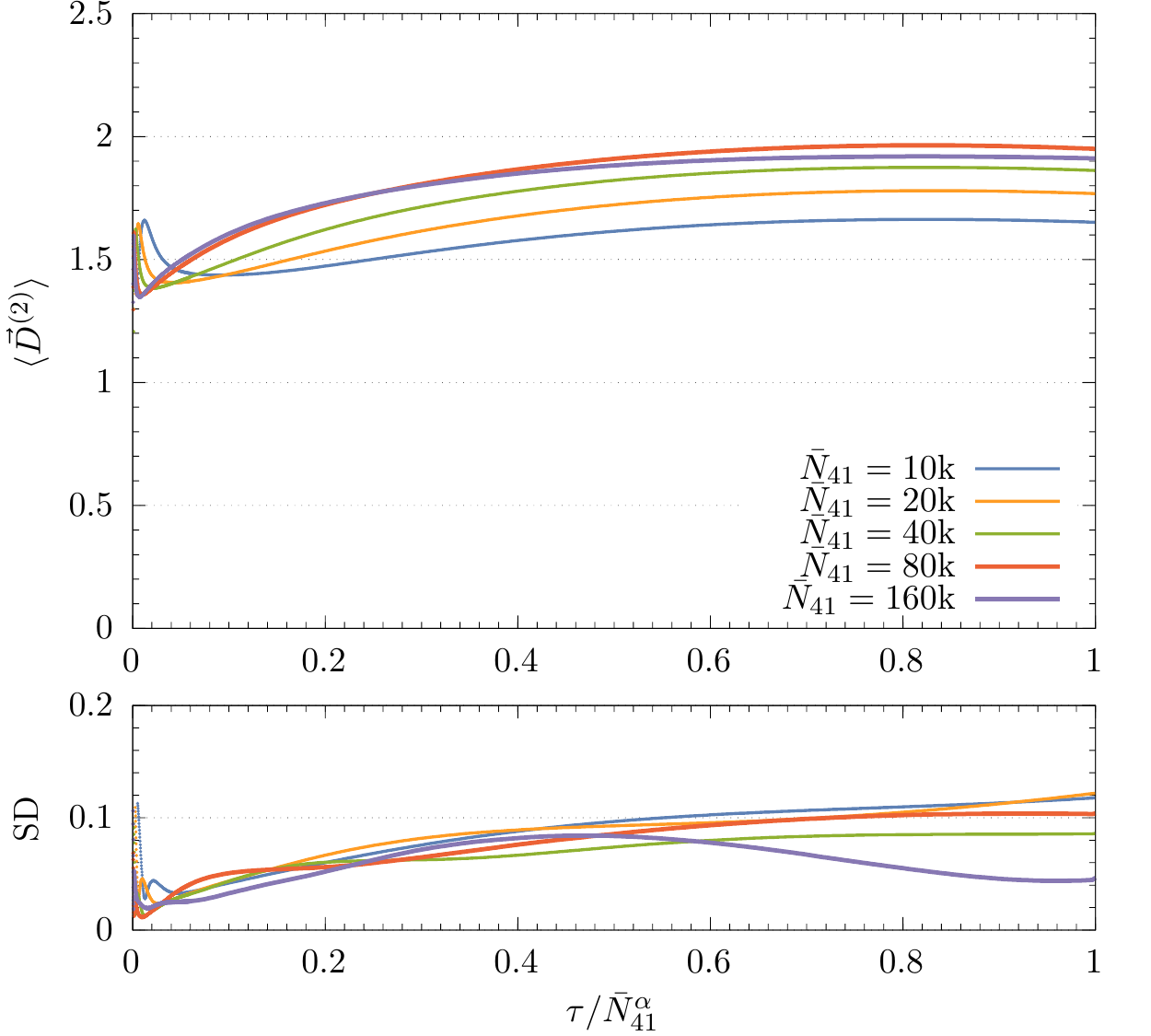}
\includegraphics[width=0.49\textwidth]{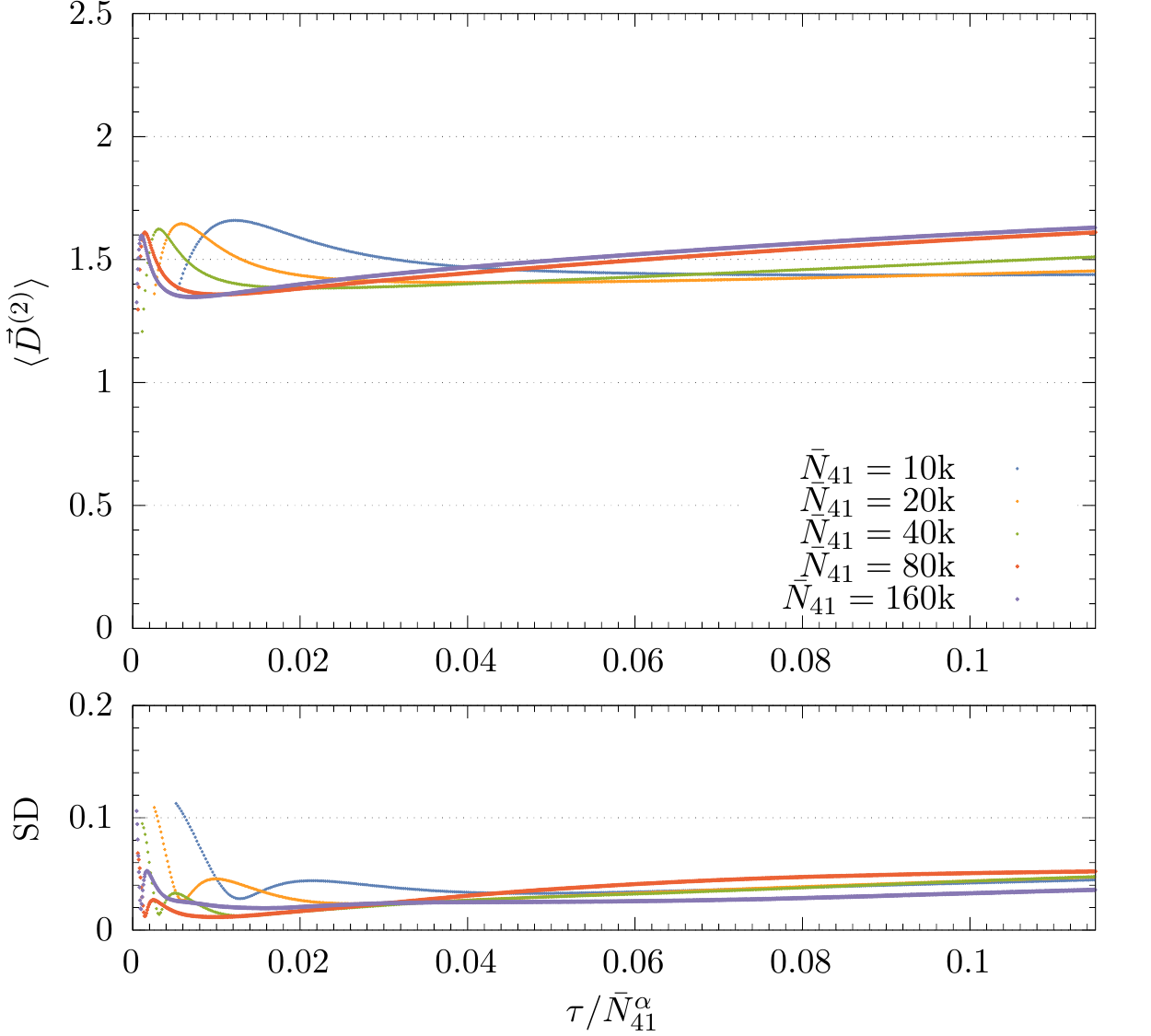}
\caption{The upper left figure shows a sequence of curves with volume-normalised diffusion time $\tau/N^{\alpha}$ of the expectation value of the tensor spectral dimension $\langle \vec{D}^{(2)} \rangle$ for geometries of different volumes. The volume of a CDT geometry is determined by the number of four-simplices, which is approximately equal to $\bar{N}_{41}$. The two curves for $\bar{N}_{41} = 160\rk, 240\rk$ are shown with a bold line. The lower left figure shows the standard deviation $SD$ of the curves in the upper left plot. The upper right figure shows $\langle \vec{D}^{(2)} \rangle$ for small diffusion times, and the lower right figure shows the corresponding standard deviation. The figures to the right show in detail how the lattice artefact peaks move to the left for increasing $\bar{N}_{41}$ and how the small diffusion time plateau emerges. The curves are shown for $\alpha=1.25$.}
\label{TriangleV}
\end{figure}

\begin{figure}[H]
\centering
\includegraphics[width=0.49\textwidth]{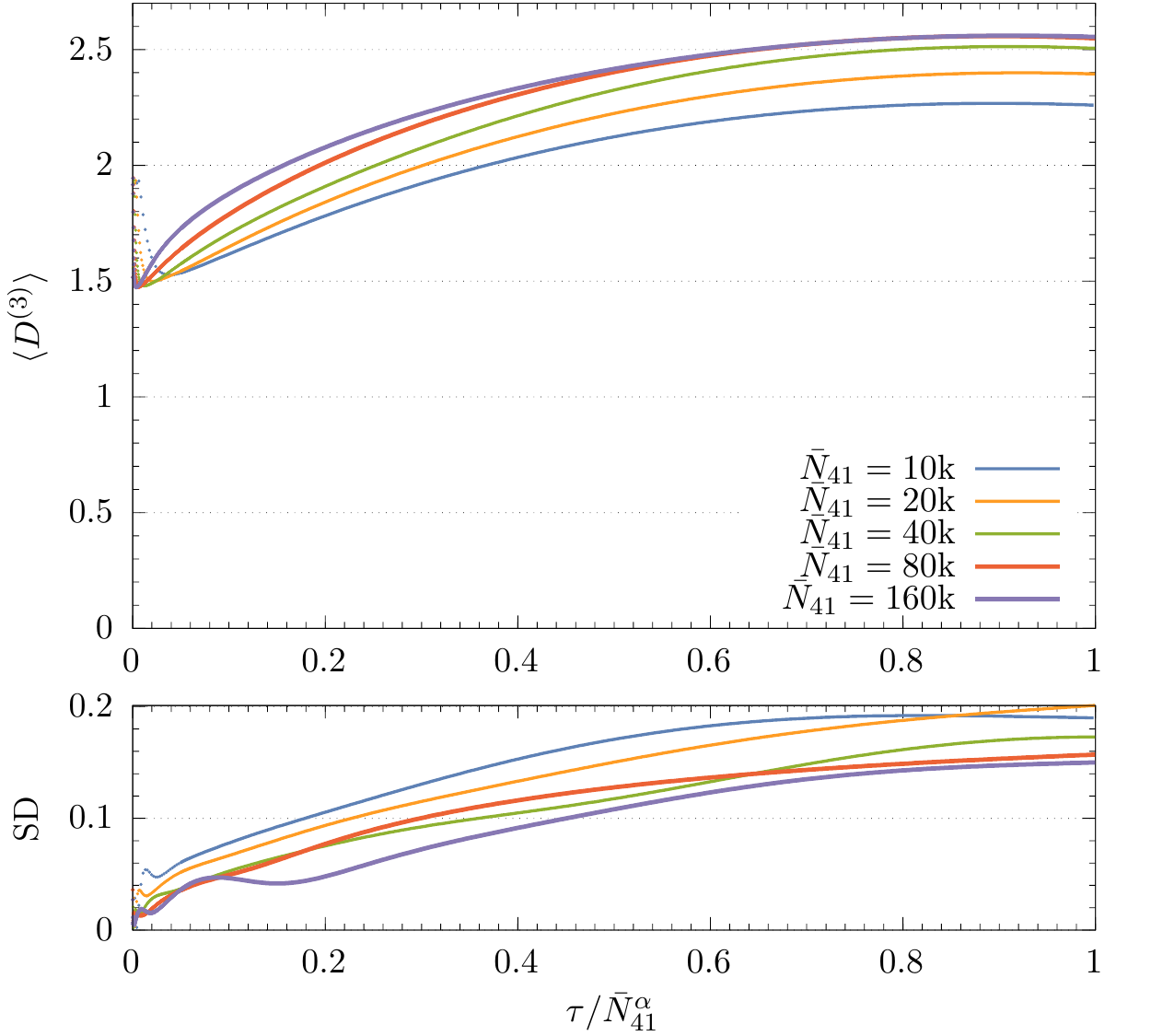}
\includegraphics[width=0.49\textwidth]{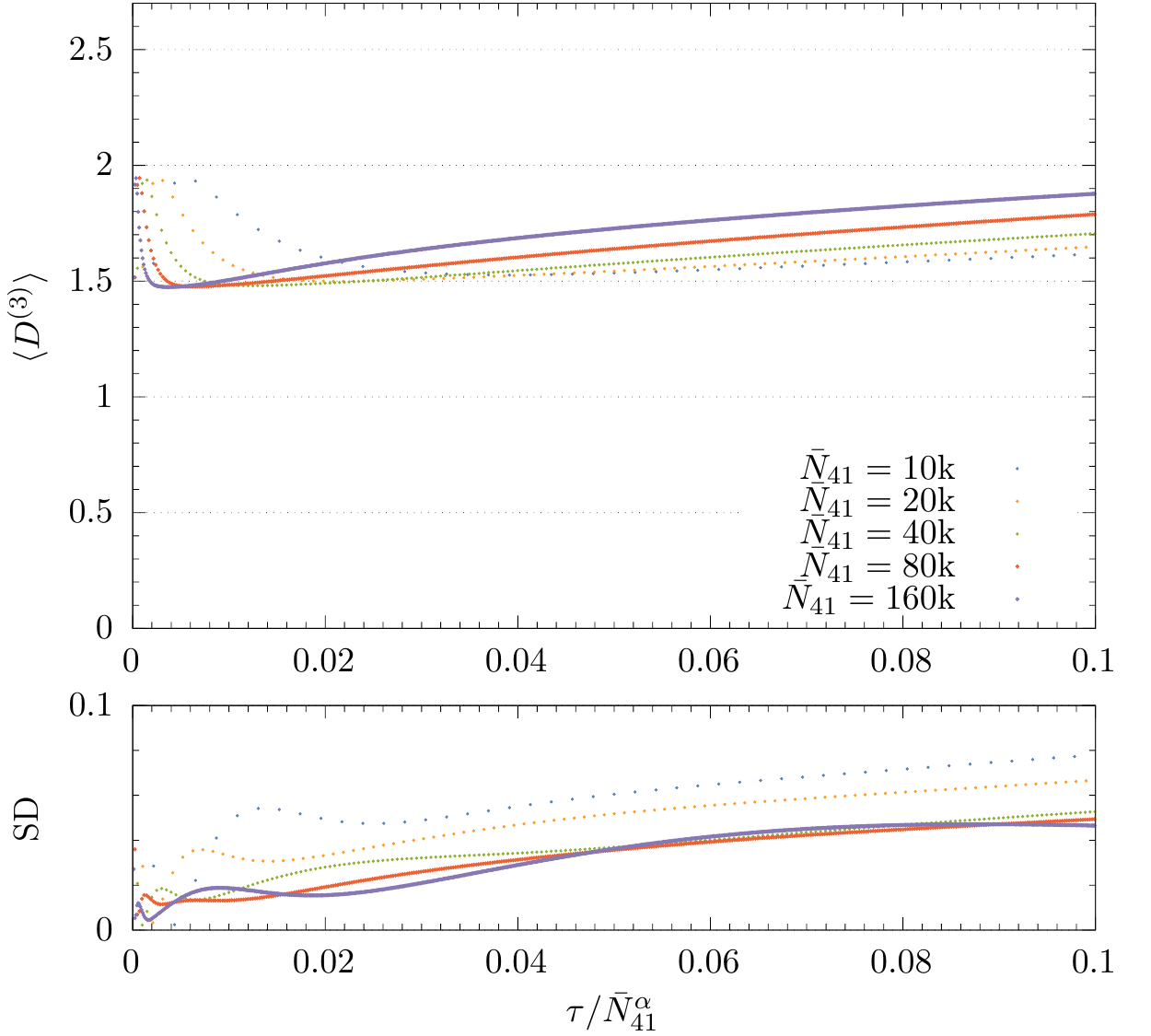}
\caption{The upper left figure shows a sequence of curves with volume-normalised diffusion time $\tau/N^{\alpha}$ of the expectation value of the dual scalar spectral dimension $\langle D^{(3)} \rangle$ for geometries of different volumes. The volume of a CDT geometry is determined by the number of four-simplices, which is approximately equal to $\bar{N}_{41}$. The two curves for $\bar{N}_{41} = 160\rk, 240\rk$ are shown with a bold line. The lower left figure shows the standard deviation $SD$ of the curves in the upper left plot. The upper right figure shows $\langle D^{(3)} \rangle$ for small diffusion times and the lower right figure shows the corresponding standard deviation. The figures to the right show in detail how the lattice artefact peaks move to the left for increasing $\bar{N}_{41}$ and how the small diffusion time plateau emerges. The curves are shown for $\alpha=1.06$.}
\label{Tetra}
\end{figure}

For the present case of CDT with toroidal topology, distinguishing between the spatial slices of the geometries has no physical significance. We have, therefore, estimated the expectation value of the average return probabilities $\langle P_i^{(k)}(\tau) \rangle$ and $\langle \vec{P}_i^{(k)}(\tau) \rangle$ for target volume $\bar{N}_{41}$ by averaging over starting points and configurations and also over slices. From $\langle P_i^{(k)}(\tau) \rangle$ and $\langle \vec{P}_i^{(k)}(\tau) \rangle$ we calculate the running spectral dimensions $\langle D^{(k)}(\tau)\rangle$ and $\langle \vec{D}^{(k)}(\tau)\rangle$ with,
\begin{equation}
\langle D^{(k)}(\tau)\rangle=- 2\tau \log(\langle P^{(k)}(\tau+s)\rangle/\langle P^{(k)}(\tau)\rangle),
\end{equation}
and
\begin{equation}
\langle \vec{D}^{(k)}(\tau)\rangle=- 2\tau \log(\langle\vec{P}^{(k)}(\tau+s)\rangle/\langle\vec{P}^{(k)}(\tau)\rangle).
\end{equation}
We also keep track of the standard deviation of the spectral dimensions, which include both statistical (from the average over starting points) as well as quantum fluctuations (from the average over geometries). We show the spectral dimensions for the scalar (Fig. \ref{Points}), one-form (Fig. \ref{Links}), vector (Fig. \ref{LinksV}), two-form (Fig. \ref{Triangle}), tensor (Fig. \ref{TriangleV}) and dual scalar diffusion (Fig. \ref{Tetra}) for different target volumes $\bar{N}_{41}$.

To be able to compare the spectral dimension curves for different target volumes, the diffusion time of the curves is normalised by $\bar{N}_{41}^\alpha$, where $\bar{N}_{41}$ is the target volume discussed in Sec. \ref{CDT}. The number of four-simplices in a CDT geometry is approximately equal to $\bar{N}_{41}$ and the average number of tetrahedra in a spatial slice is proportional to $\bar{N}_{41}$.

The exponent $\alpha$ is chosen such that the maximum of the curves for large diffusion time are aligned for the same volume-normalised diffusion time $\tau/\bar{N}_{41}^{\alpha}$. As was shown in equation \eqref{GapDom}, the exponent $\alpha$ for a differentiable manifold $\mathcal{M}$ is related to the dimension $n$ of $\mathcal{M}$. Although we could therefore potentially define an alternative definition of an effective dimension through $\alpha$, we will leave a detailed discussion of $\alpha$ for a future investigation. Each of the spectral dimension curves are shown up to a time-scale at which the zero modes become dominant and the curves monotonically go to zero. We find that all spectral dimensions except $D^{(0)}$ exhibit a clear finite-volume scaling, which signals the existence of an infinite volume limit for the spectral dimension curves. Plateaus in the infinite volume curve signal the appearance of an effective dimension for some range of scales, as was illustrated with equation \eqref{SpecDimForm} and the examples of regular geometries in Figs. \ref{dfig:3proof-of-concept} and \ref{FlatDims}. The examples of regular geometries showed that plateaus take the form of an extremum or inflection point of the spectral dimension curve surrounded by a (small) range of diffusion time where the spectral dimension changes slowly. From the finite volume scaling of the maximum of the plateaus, we can extrapolate an effective dimension. The extrapolated values of the effective dimensions derived from $\langle D^{(k)} \rangle$ and $\langle \vec{D}^{(k)} \rangle$ are summarised in Tables~\ref{table:1} and~\ref{table:2}, which can be found at the end of this section.

For the scalar spectral dimension $D^{(0)}$ (Fig. \ref{Points}), we see that a plateau appears for larger volumes similar to those of the regular geometries discussed in Sec. \ref{regular}. The initial peak of the curves shrinks for larger volumes and was also present for the regular geometries as a spurious effect. We therefore conclude that the initial peak most likely is a lattice artefact. However, for the target volumes $\bar{N}_{41}$ we have considered, the initial peak and the plateau cannot be disentangled and we cannot extrapolate an infinite-volume value for the plateau. From the maximum of the plateau of $\langle D^{(0)} \rangle$ for the largest volume we could expect that an infinite volume limit will be close to four. However, it remains to be seen if including geometries of larger volume indeed makes it possible to extrapolate an infinite volume limit more accurately. We also note that there is no other timescale at which $\langle D^{(0)} \rangle$ exhibits a flattening of the curve. The curves of $\langle D^{(0)} \rangle$ start at zero and, for larger volumes, are increasingly steeper for small diffusion times.

Both the expectation value of the one-form spectral dimension $\langle D^{(1)} \rangle$ (Fig. \ref{Links}), and of the vector spectral dimension $\langle \vec{D}^{(1)} \rangle$ (Fig. \ref{LinksV}), show an increasingly steep initial increase from zero to a plateau, implying that there exists no small-scale spectral dimension. The plateau is more elongated for the vector spectral dimension, indicating that finite-size effects are less dominant than for the case of the one-form spectral dimension. We can extrapolate an effective dimension for both $\langle D^{(1)} \rangle$ and $\langle \vec{D}^{(1)} \rangle$ from fitting the finite volume scaling of the maximum of the emerging plateaus to a power law of the form $a+b*\bar{N}_{41}^c$.

\begin{figure}[H]
\centering
\includegraphics[width=0.49\textwidth]{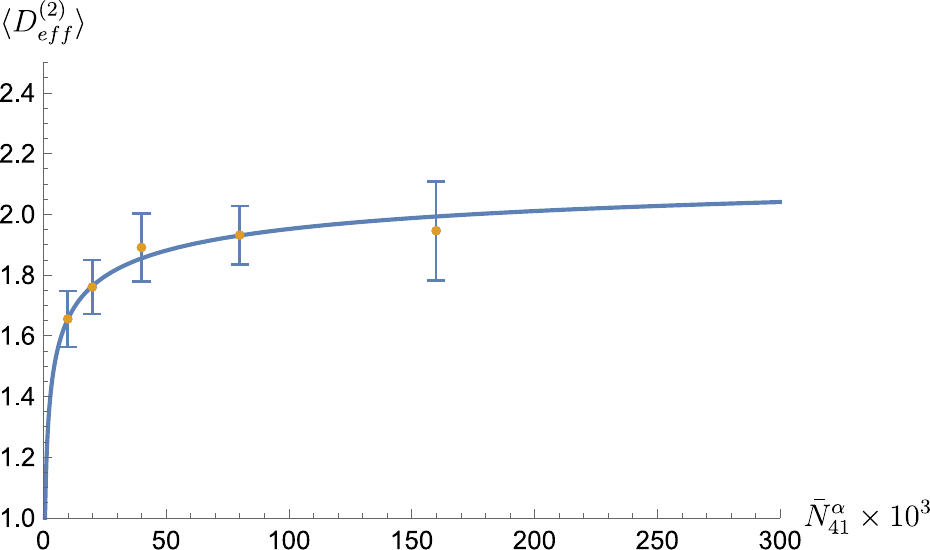}
\includegraphics[width=0.49\textwidth]{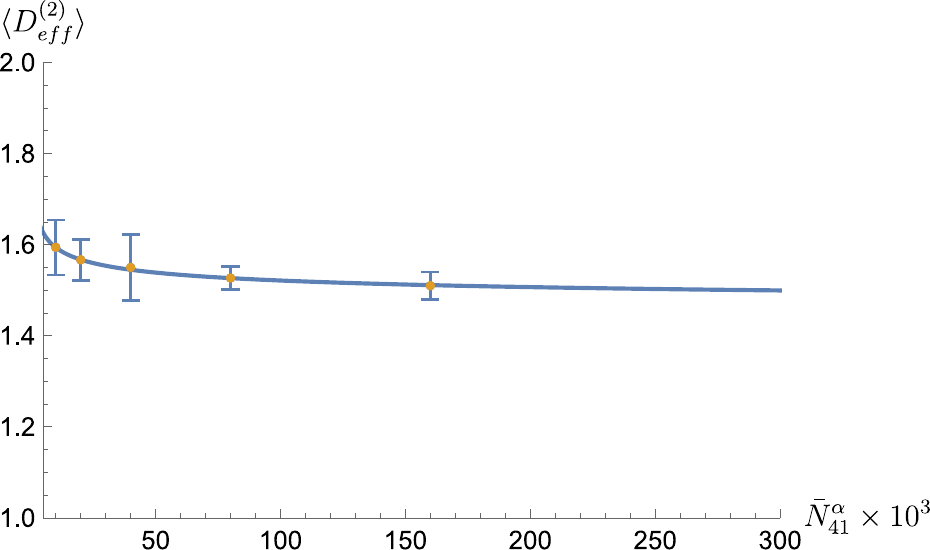}
\caption{The left figure shows the finite volume scaling of the maxima of the emerging plateaus of $\langle D^{(2)} \rangle$ at large diffusion time as a function of $\bar{N}_{41}$. The same plot shows a power law fit of the form $a+b*\bar{N}_{41}^{ c}$, from which we extrapolate an IR effective dimension $\langle D_{eff}^{(2)}\rangle = 2.30 \pm 0.64$ in the infinite volume limit. The right figure shows the finite volume scaling of the minima of the emerging plateaus of $\langle D^{(2)} \rangle$ at small diffusion time. From the corresponding power law fit we extrapolate a UV effective dimension $\langle D_{eff}^{(2)} \rangle = 1.44 \pm 0.23$. }
\label{Triangleeff}
\end{figure}

\begin{figure}[H]
\centering
\includegraphics[width=0.49\textwidth]{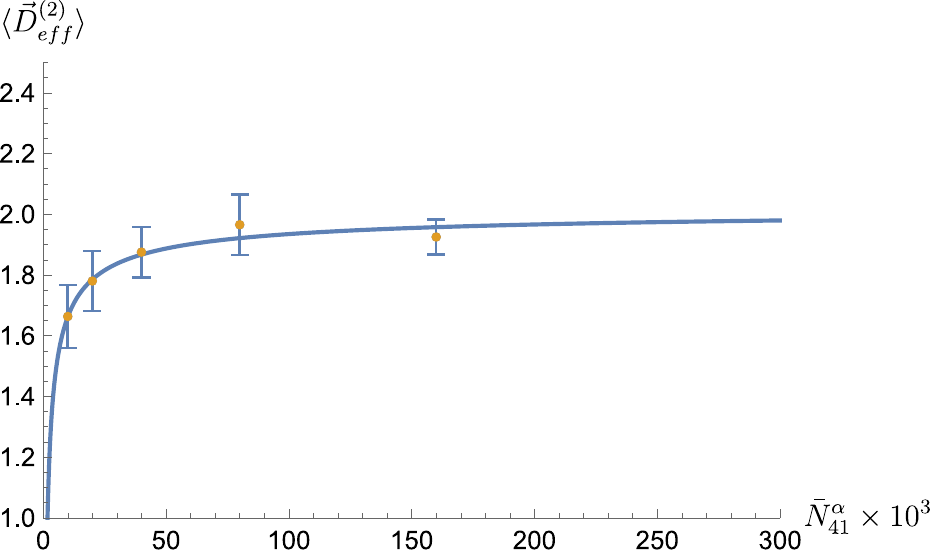}
\includegraphics[width=0.49\textwidth]{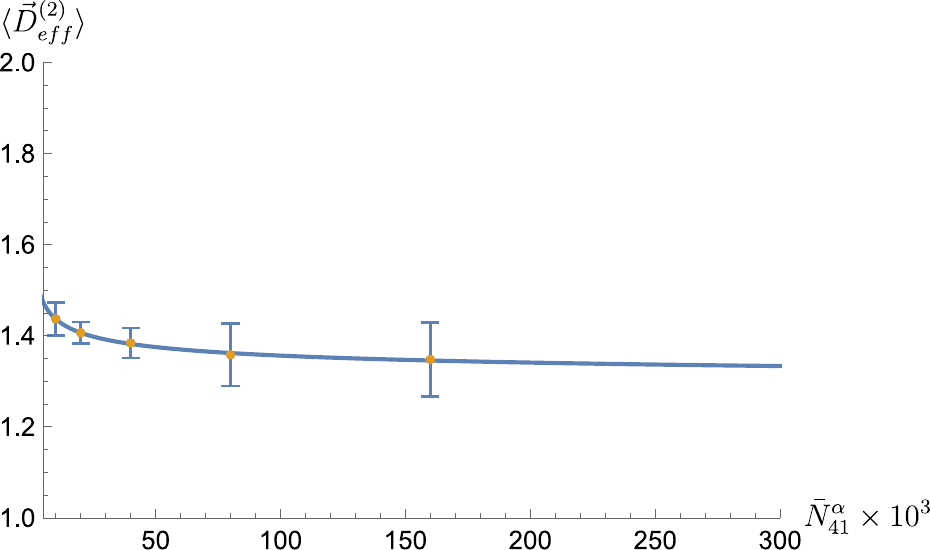}
\caption{The left figure shows the finite volume scaling of the maxima of the emerging plateaus of $\langle \vec{D}^{(2)} \rangle$ at large diffusion time as a function of $\bar{N}_{41}$. The same plot shows a power law fit of the form $a+b*\bar{N}_{41}^c$, from which we extrapolate an IR effective dimension $\langle \vec{D}_{eff}^{(2)} \rangle = 2.03 \pm 0.11$ in the infinite volume limit. The right figure shows the finite volume scaling of the minima of the emerging plateaus of $\langle \vec{D}^{(2)} \rangle$ at small diffusion time. From the corresponding power-law fit we extrapolate a UV effective dimension $\langle \vec{D}_{eff}^{(2)} \rangle = 1.27 \pm 0.26$. }
\label{TriangleVeff}
\end{figure}

The two-form spectral dimension $\langle D^{(2)} \rangle$ (Fig. \ref{Triangle}) and the tensor spectral dimension $\langle \vec{D}^{(2)} \rangle$ (Fig. \ref{TriangleV}) have more structure than the scalar, one-form and vector spectral dimensions. For small diffusion times $\langle D^{(2)} \rangle$ and $\langle \vec{D}^{(2)} \rangle$ exhibit a sharp peak that moves to $\tau/N^\alpha=0$ for larger volumes. The volume scaling behaviour of the peak at small diffusion times indicates that these peaks are lattice artefacts. Immediately after the peak the curves of $\langle D^{(2)} \rangle$ and $\langle \vec{D}^{(2)} \rangle$ show a plateau at small diffusion times with a clear finite volume scaling. We therefore conclude that the infinite volume limit exhibits an effective dimension in the UV for both the two-form and the tensor case. The values of the extrapolated effective dimension for $\langle D^{(2)} \rangle$ and $\langle \vec{D}^{(2)} \rangle$ agree within one standard deviation. For significantly larger diffusion times, a second plateau appears for both $\langle D^{(2)} \rangle$ and $\langle \vec{D}^{(2)} \rangle$ for which the extrapolated effective dimensions also agree within one standard deviation. The finite volume scaling and extrapolated effective dimension of the UV and IR plateaus are given in Fig. \ref{Triangleeff} for $\langle D^{(2)} \rangle$ and in Fig. \ref{TriangleVeff} for $\langle \vec{D}^{(2)} \rangle$.

\begin{figure}[H]
\centering
\includegraphics[width=0.49\textwidth]{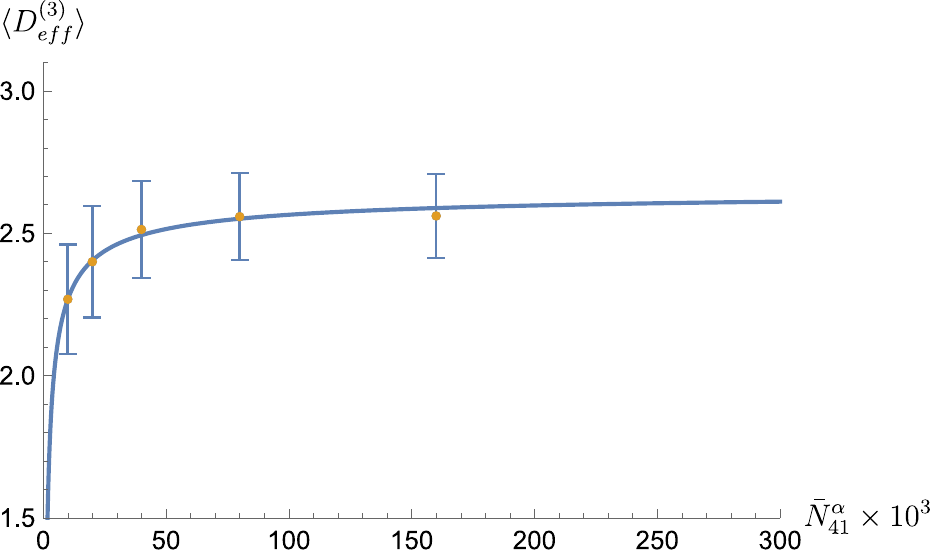}
\includegraphics[width=0.49\textwidth]{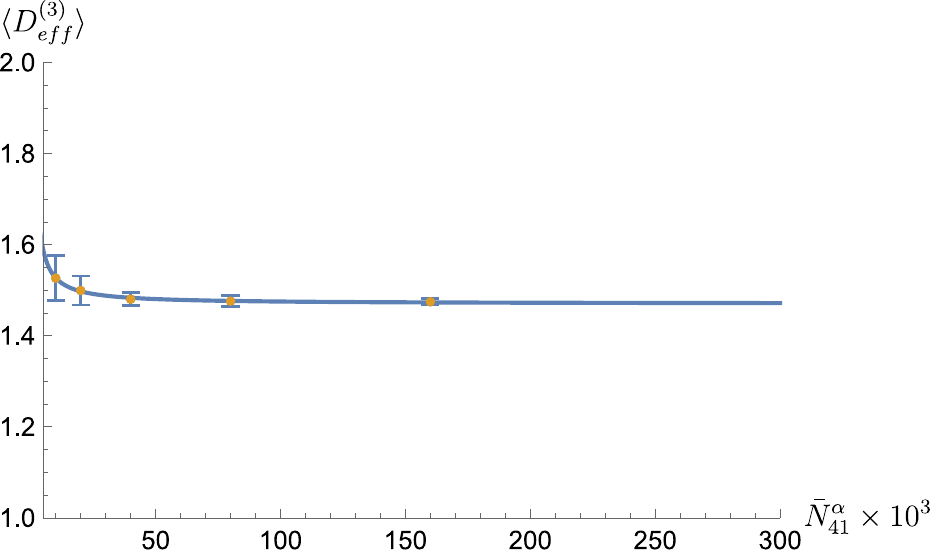}
\caption{The left figure shows the finite volume scaling of the maxima of the emerging plateaus of $\langle D^{(3)} \rangle$ at large diffusion time as a function of $\bar{N}_{41}$. The same plot shows a power law fit of the form $a+b*\bar{N}_{41}^c$, from which we extrapolate an IR effective dimension $\langle D_{eff}^{(3)} \rangle = 2.66 \pm 0.21$ in the infinite volume limit. The right figure shows the finite volume scaling of the minima of the emerging plateaus of $\langle D^{(3)} \rangle$ at small diffusion time. From the corresponding power law fit we extrapolate a UV effective dimension $\langle D_{eff}^{(3)} \rangle = 1.47 \pm 0.01$. }
\label{Tetraeff}
\end{figure}

The dual scalar spectral dimensions $\langle D^{(3)} \rangle$ (Fig. \ref{Tetra}), like the two-form and tensor spectral dimension, exhibits a sharp initial peak that moves to $\tau/N^\alpha=0$ for larger volumes. Therefore, this initial peak is interpreted as a lattice artefact as well. The subsequent plateau at small diffusion times persists for large volumes and shows a clear finite volume scaling so signals the presence of a UV effective dimension. For significantly larger diffusion times a second plateau emerges which exhibits a finite volume scaling as well. We can therefore also extrapolate an IR effective dimension for $\langle D^{(3)} \rangle$ (Fig. \ref{Tetraeff}). A summary of the results for the UV and IR spectral dimensions we found for all $\langle D^{(k)} \rangle$ and $\langle \vec{D}^{(k)} \rangle$ are given in Tables \ref{table:1} and \ref{table:2}.

\begin{table}[htb!]
\centering
\begin{tabular}{||c c||} 
 \hline
 Type & Effective Spectral Dimension  \\ 
 \hline\hline
 Scalar $\langle D^{(0)} \rangle$ & $\sim 4$ \\ 
 One-form $\langle D^{(1)} \rangle$ & $4.98 \pm 0.36$  \\
 Vector $\langle \vec{D}^{(1)} \rangle$ & $4.03 \pm 0.45$ \\
 \hline
\end{tabular}
\caption{The table contains the extrapolated infinite volume values of the effective spectral dimension related to the scalar, one-form, and vector diffusion.}
\label{table:1}
\end{table}

\begin{table}[htb!]
\centering
\begin{tabular}{||c c c||} 
 \hline
 Type & UV & IR  \\ 
 \hline\hline
 Two-form $\langle D^{(2)} \rangle$& $1.44 \pm 0.23$ & $2.30 \pm 0.64$ \\
 Tensor $\langle \vec{D}^{(2)} \rangle$ & $1.27 \pm 0.26$ & $2.03 \pm 0.11$ \\  
 Dual scalar $\langle D^{(3)} \rangle$& $1.47 \pm 0.01$ & $2.66 \pm 0.21$ \\  
 \hline
\end{tabular}
\caption{The table contains the extrapolated close range (UV) and long range (IR) values of the effective spectral dimensions related to the two-form, tensor and dual scalar diffusion.}
\label{table:2}
\end{table}



\subsection{Spatial Hausdorff dimension}

\begin{figure}[h]
    \centering
    \includegraphics[width = 0.8\textwidth]{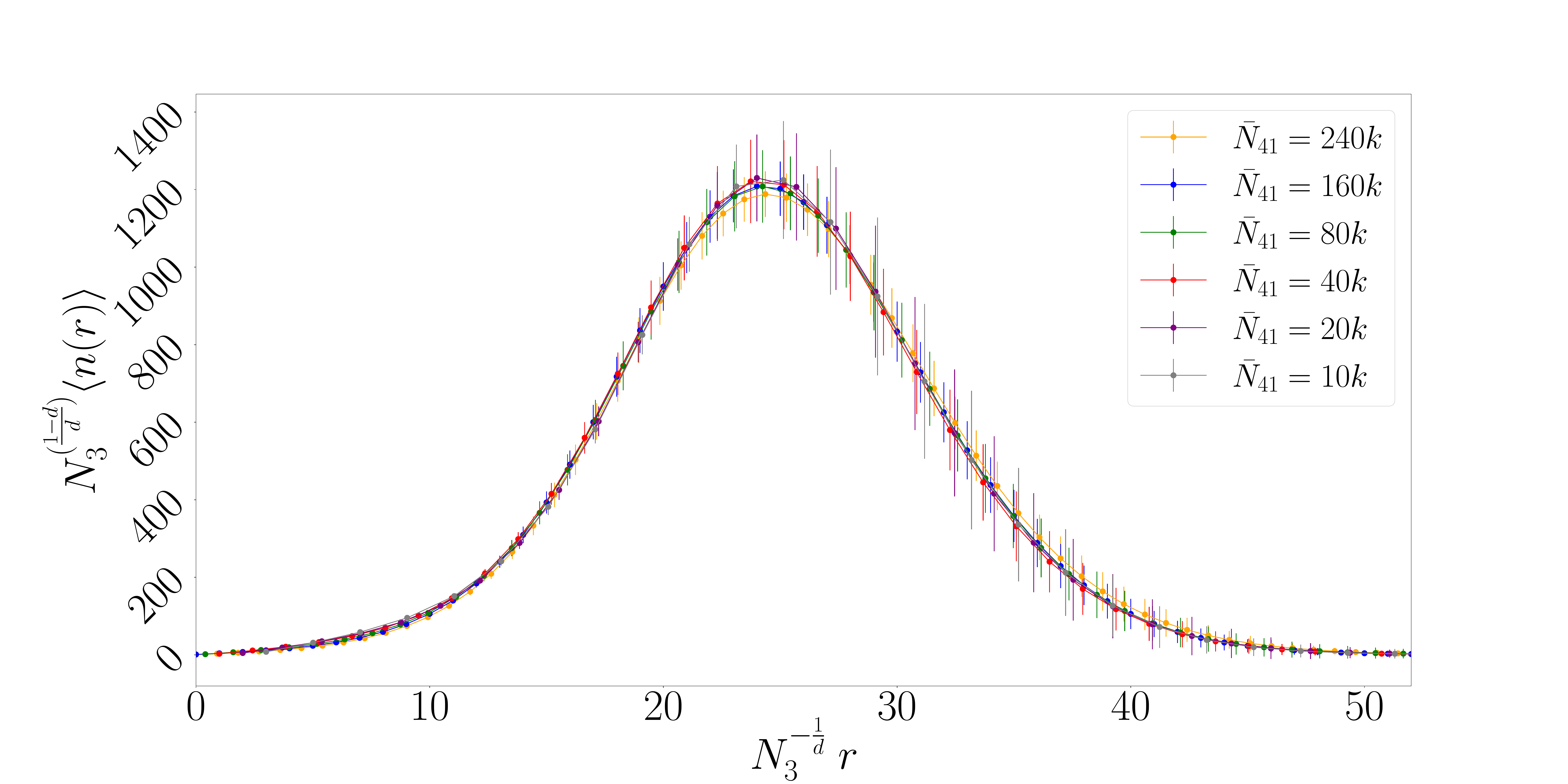}
    \caption{ The figure shows the expectation values of the shell volumes $N_3^{\frac{1-d}{d}} \langle n(r) \rangle$ as a function of radius $N_3^{-\frac{1}{d}} r$ rescaled with respect to the Hausdorff dimension $d$. The plots include error bars and different colours correspond to different target volumes $\bar{N}_{41}$. The best overlap of rescaled shell volumes for different target volumes is obtained for $d_H = 3.95 \pm 0.18 $.}
    \label{fig:hausdorff}
\end{figure}

As a point of comparison for the results on the various spectral dimensions, we have also considered another observable related to the effective dimension of spacetime, the Hausdorff dimension, which has been studied extensively in the context of non-perturbative quantum gravity \cite{cdt_desit_fi}. The Hausdorff dimension for a geometry $T$ is defined through the scaling of the volume $V(r)$ of balls with the radius $r$. We use the tetrahedron adjacency matrix of each spatial slice of $T$ to construct balls consisting of all tetrahedra within the lattice distance $r$ from a central tetrahedron. Tetrahedra at a geodesic distance $r$ from a central tetrahedron form discrete shells of these balls. A shell volume $n(r)$ denotes the number of tetrahedra in a shell of radius $r$, the total volume of all shells constitutes the number of tetrahedra in a slice $N_3 = \sum_{r} n(r)$. The average of $n(r)$ over centre points, slices and geometries $T$ gives the expectation value $\langle n(r) \rangle$ of shell volumes. The Hausdorff dimension can be obtained from the finite volume scaling of $n(r)$. In order to compare shell volume profiles $n(r)$ for various slice volumes $N_3$, we rescale the radius $r$ to $N_3^{-\frac{1}{d}} r$
with respect to the Hausdorff dimension $d$ and normalise the shell volumes $n(r)$ to $N_3^{\frac{1-d}{d}} n(r)$, so that an integral over $r$ of rescaled shell volumes is equal to $1$. If there is a common Hausdorff dimension $d = d_H$ for geometries of different sizes, the rescaled curves will overlap. Fig. \ref{fig:hausdorff} shows the rescaled average curves calculated for the for the same CDT geometries we have studied for the spectral dimensions $\langle D^{(k)} \rangle$ and $\langle \vec{D}^{(k)} \rangle$. The best fit for $d$ results in a Hausdorff dimension $d_H = 3.95 \pm 0.18 $, which is different than $3$, the dimension of the tetrahedra that make up the slices.

\section{Discussion and outlook}
\label{Discussion}
The aim of this work was to investigate whether the universal phenomenon of scale-dependent effective dimensions in non-perturbative theories of quantum gravity depends on the type of field that probes the effective spacetime. To investigate this question, we have studied the scaling of the trace of the heat kernel of the Laplace-Beltrami operator defined on $k$-form fields on spatial slices of the non-perturbative model of quantum gravity known as Causal Dynamical Triangulations. From the scaling of the trace of the heat kernel, we can derive the scale-dependent spectral dimension, which was extensively studied in various theories of non-perturbative quantum gravity for the cases $k=0$ and $k=n$. To our knowledge, this work is the first time that cases $k \in \{1,...,n-1\}$ have been studied. As mentioned in the introduction, one of the important tasks for quantum gravity research is to find examples of quantum gravity phenomenology, especially in view of the recent technological advances in gravitational wave observations and direct imaging of black hole shadows. It has been argued, using a model-independent framework, that the seemingly universal scale-dependent effective dimension can have an effect on the relative propagation of a gravitational wave signal and the corresponding electromagnetic signal \cite{Calcagni1, Calcagni2, Calcagni3}. However, it was assumed that only the scalar mode of the gravitational waves were affected and that the propagation of the electromagnetic signal was unaffected by the running effective dimension. This study of the various spectral dimensions can be a starting point to investigate these assumptions in a non-perturbative theory of quantum gravity. We have found that the two-form, tensor and dual scalar spectral dimensions exhibit two scales at which an effective dimension appears, i.e. the curve has two regimes with an extremum. However, the one-form and vector spectral dimension show only a single scale at which an effective dimension emerges, i.e. the curve only shows one extremum. The fact that the one-form and vector spectral dimensions do not show a dimensional flow between two scales can potentially be related to the absence of a dispersion relation for the electromagnetic field, but dynamically generated instead of as an ad hoc assumption. It is, however, as of yet unclear how to directly relate scale-dependent effective dimensions to dispersion relations for the propagation of physical fields.

Before an attempt can be made to answer an ambitious phenomenological question such as the one stated above, it will be advantageous to better understand the dynamics of the various spectral dimensions. For example, it is clear that the size and number of the geometries studied was too small to extrapolate an infinite volume limit from the scalar spectral dimension $\langle D^{(0)}\rangle$, because the lattice artefacts in $\langle D^{(0)}\rangle$ were strong in the measurements we made. In the simplicial geometries under consideration, the number of vertices is small in comparison to the simplices of higher dimension, so it is expected that lattice artefacts are harder to control.

For small diffusion times, $\langle D^{(3)} \rangle$ shows a local minimum after an initial peak that we consider to be a lattice artefact, just like the initial peaks for the regular geometries in Figs. \ref{dfig:3proof-of-concept} and \ref{FlatDims}. For these regular geometries, we know that the initial peaks are lattice artefacts, because we know the dimension of the manifolds the regular geometries approximate. We have identified the local minimum of $\langle D^{(3)} \rangle$ as the UV plateau of $\langle D^{(3)} \rangle$. The value of the UV plateau is at approximately $1.5$, which is in agreement with the value of the same spectral dimension obtained for spatial slices of four-dimensional CDT with spherical spatial topology \cite{SpecDimCDT2}. We note here that the plateau of $\langle D^{(3)} \rangle$ in \cite{SpecDimCDT2} was obtained for the largest diffusion time of that numerical study, while $\langle D^{(3)} \rangle$ in Fig. \ref{Tetra} takes the value $1.5$ for small diffusion time. The size of the geometries and the maximal diffusion time of the numerical study in \cite{SpecDimCDT2} was however much smaller (between $500$ and $2000$ tetrahedra), because of the computer resources that were available at that time. The value obtained for the long range diffusion time of $\langle D^{(3)} \rangle$ in \cite{SpecDimCDT2} therefore should be compared to the short range (up to approximately $\tau/\bar{N}_{41}=0.2$) value of $\langle D^{(3)} \rangle$ of the $20k$ and $10k$ curves\footnote{A spatial slice of a typical geometry with $\bar{N}_{41}=20k$ with $4$ slices consists of approximately $2500$ tetrahedra, because every tetrahedron is shared by two four-dimensional simplices. The diffusion time of $200$ discrete steps corresponds to approximately $\tau/\bar{N}_{41}=0.2$ in Fig. \ref{Tetra} of the $20k$ curve.} in Fig. \ref{Tetra}. Reobtaining the value $1.5$ for $\langle D^{(3)} \rangle$ is an indication that results for the effective dimension at small scales are universal, even though the choice of normalisation\footnote{The choice of normalisation was discussed in Sec. \ref{Discretuum}.} of the Laplacian $L^{(3)}$ and the topology of the geometries in the current work is different than the normalisation and topology in \cite{SpecDimCDT2}. An IR spectral dimension has not been previously reported for spatial slices in CDT. Our measurements were performed at a specific point $(\kappa_0=4.0,\Delta= 0.2)$ in the CDT phase diagram, at which we expect a much smaller effective lattice spacing than at the point $(\kappa_0=2.2,\Delta= 0.6)$ in phase $C$, where most previous spectral dimension measurements were made. From the point of view of the conjectured RG trajectories in the CDT phase diagram \cite{Ambjorn:2014gsa,Renormalization_Ambjorn2020}, we expect that the effective theory at $(\kappa_0= 4.0,\Delta= 0.2)$ captures more features of the deep quantum regime and the potential UV fixed point than the effective theory at $(\kappa_0= 2.2,\Delta= 0.6)$.

In a (semi-)classical regime, the various spectral dimensions should be (approximately) equal for large diffusion times, as is the case for a differentiable manifold. As presented in Tab. \ref{table:2}, our results show a potential agreement, within our measurement accuracy, between the IR values of the effective dimension of two-form $\langle D^{(2)}\rangle$, tensor $\langle \vec D^{(2)}\rangle$ and dual scalar $\langle D^{(3)}\rangle$. The UV values ($\approx 1.5 $) of $\langle D^{(2)}\rangle$, $\langle \vec D^{(2)}\rangle$ and $\langle D^{(3)}\rangle$ agree within error bars as well. On the contrary, we observe completely different behaviour for the one-form $\langle D^{(1)}\rangle$ and vector $\langle \vec D^{(1)}\rangle$ spectral dimensions (Tab. \ref{table:1}), which show only a single effective scale. We also note that only the vector effective dimension agrees with the Hausdorff dimension ($d_H=4$) and that none of the spectral dimensions is consistent with the topological dimension of the spatial slices, which is equal to $3$. All in all, we have established that the effective geometry of CDT in the deep quantum regime exhibits a dynamically generated effective dimension that shows significantly qualitative differences for different $k$-form fields. Exact numerical values in the UV regime are however difficult to establish due to lattice and discretisation artefacts. We also conclude that the ensembles of spatial slices we have investigated do not represent an effective geometry which is close to a classical manifold. This is to be expected as previous studies of spatial slices also showed that their geometry is non-classical \cite{SpecDimCDT2, AMBJORN2010420}.

It is difficult to compare the physical scales at which the effective dimensions appear for different $k$. One reason for this is that the discrete Laplace-Beltrami operator, constructed from the unweighted incidence matrices of the $k$-simplices, is not defined in a manner consistent between different values of $k$. The fraction of the geometry that is probed with a discrete diffusion step from some starting point is therefore determined by the connectivity matrix for the $k$-simplex. It is non-trivial how to relate the connectivity matrices for different $k$, so it is also difficult to relate the diffusion time for different $k$. It will be interesting to explore to which extent an alternative choice of weights for the incidence matrix makes it possible to compare scales at which effective dimensions appear for different values of $k$. A potential candidate for such a choice of weights is given by the framework of Discrete Exterior Calculus (DEC) \cite{Desbrun}. Using DEC, we can choose weights for the incidence matrices such that a diffusion step is weighted by the associated dual volume of the $k$-simplices.

Another interesting avenue to explore would be to investigate all spectral dimensions for a sequence of points in the phase diagram. Such a study can indicate if the RG flows towards the IR of the effective dimensions for large diffusion time converge. If we find such an RG trajectory, we can also potentially answer the question whether the $k$-form spectral dimension $\langle D^{(k)} \rangle$ or the $k$-tensor spectral dimension $\langle \vec{D}^{(k)} \rangle$ better captures the physical properties of the effective dimension. In the present study $\langle D^{(k)} \rangle$ and $\langle \vec{D}^{(k)} \rangle$ clearly have different properties, while they are approximately equal for the regular geometries. We found that lattice artefacts for small diffusion time are weaker for $\langle \vec{D}^{(k)} \rangle$ and that the corresponding $k$-tensor fields have faster convergence properties. However, these properties are not sufficient to rule out the relevance of the effective dimension derived from $\langle D^{(k)} \rangle$. Therefore, we have chosen to present the results for both $\langle D^{(k)} \rangle$ and $\langle \vec{D}^{(k)} \rangle$ in this text.

It is also interesting to consider the four-dimensional geometries for which the dual scalar spectral dimension has been studied extensively in the literature. In the four-dimensional case, the dual scalar\footnote{In the case of four dimensional CDT, the dual scalar Laplacian is determined by the connectivity of the four-dimensional simplices.} spectral dimension has been shown to be equal to four for large diffusion times. Traditionally, this result has been interpreted as an indication that CDT shows some (semi-)classical properties in phase $C$. It would be interesting to see to what extent the other spectral dimensions support this interpretation. The numerical challenges are greater for the four-dimensional geometries, but are in essence unchanged from what was presented in this text. For four-dimensional geometries there will also be one additional spectral dimension, because $k$ will take values in $\{ 0,...,4 \}$. 
\newpage
\section*{Acknowledgements}
The research in this work was supported by the National Science Centre, Poland, under grant no. 2019/33/B/ST2/00589. We would also like to express our deep gratitude to Prof. Jerzy Jurkiewicz (from the Institute of Theoretical Physics of the Jagiellonian University in Krakow) who collaborated with us in this research; unfortunately he unexpectedly passed away on 30th November 2021, before we finished this work.

\appendix

\section{Analytical calculation of generalised spectral dimensions on the flat two-torus}
\label{app}

\begin{figure}[H]
    \centering
    \includegraphics[scale=0.7]{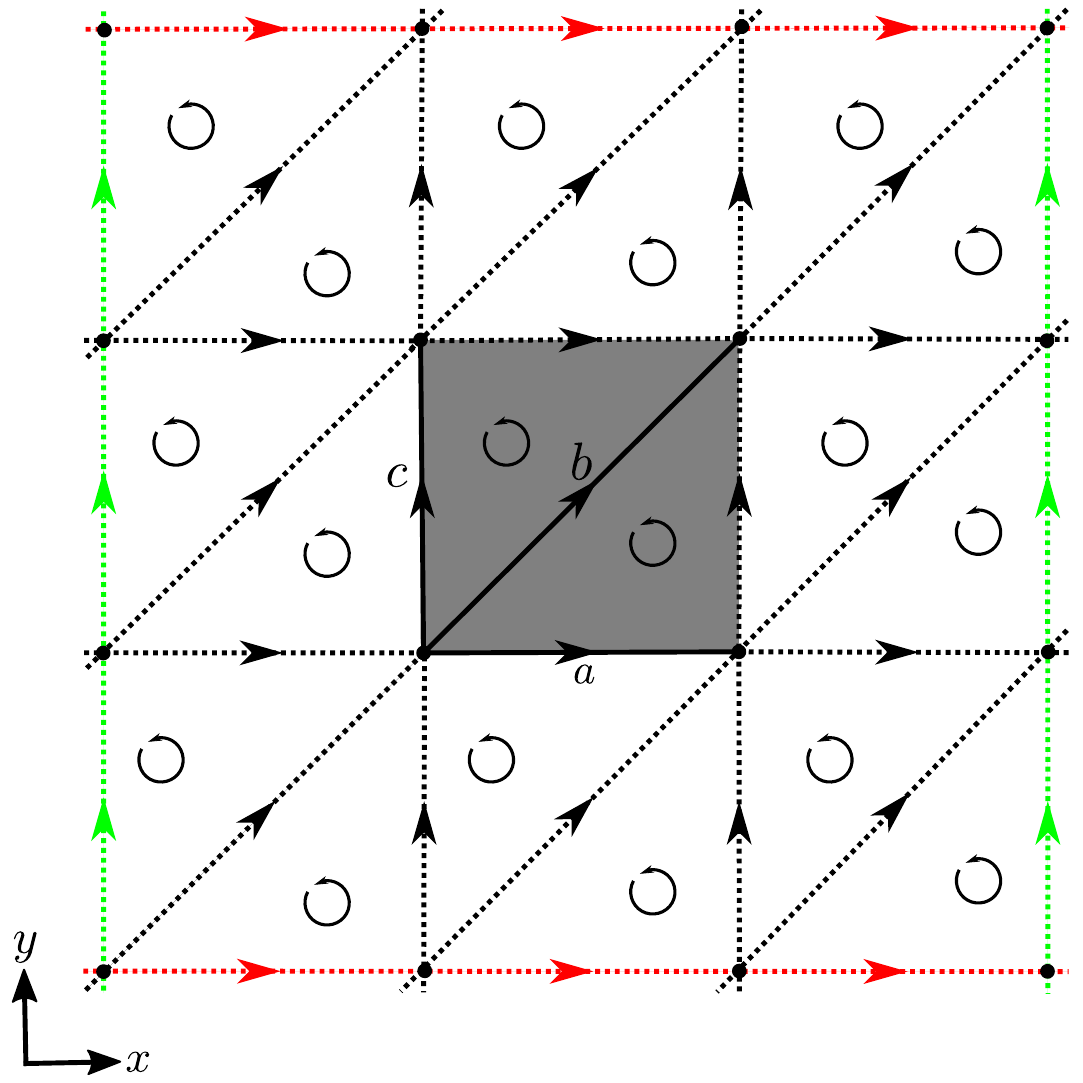}
    \caption{The figure shows a triangulation of the flat torus. The elementary cell consisting of three oriented edges is coloured gray. The red and green boundaries are identified, respectively.}
    \label{fig:d10orientededges}
\end{figure}

The following scheme describes a triangulation of the flat two-dimensional torus. We repeat the elementary square in both $x$ and $y$ direction, such that every vertex will be of degree six (see figure \ref{fig:d10orientededges}) and by pairwise identifying the sides of the triangulation we obtain a finite lattice with periods $L_x$ and $L_y$.
The link Laplacian $\Delta^{(1)}$ consists of two components as can be seen from Eq. \eqref{DiscLap}. The component of $\Delta^{(1)}$ that corresponds to the vertex-edge incidence matrix $I^1_0$ is denoted by $\mathbf{F}$ and the component of $\Delta^{(1)}$ that corresponds to the edge-triangle incidence matrix $I^2_1$ is denoted by $\mathbf{G}$. The matrices $\mathbf{F}$ and $\mathbf{G}$ can be given in the coordinate basis of the vertices parameterised by two integers ${k, l}$. We can equivalently describe the matrices $\mathbf{F}$and $\mathbf{G}$ in terms of the Fourier basis $\hat{\delta}_{kl}$ parameterised by two wave numbers ${p, q}$ such that
 \begin{align}
     \hat{\delta}_{kl}(p, q) = \mathcal{N} e^{i (kp + lq)},
 \end{align}
 where $\mathcal{N}$ is the normalisation factor. As a consequence of the periodicity, the wave numbers $p$ and $q$ take the values $p_m = \frac{2\pi}{L_x}m$ and $q_n = \frac{2\pi }{L_y}n$ where $m = 0, 1, \ldots, L_x -1$ and $n = 0, 1, \ldots, L_y - 1$.
 
 It can be shown that the Fourier transform of the matrices $\mathbf{F}$ and $\mathbf{G}$ are block-diagonal in each ${p, q}$ sector. Each block has the form
 \begin{align}
     \mathbf{F}(p, q) = \begin{pmatrix}
                            \mathbf{F}_{aa} & \mathbf{F}_{ab} & \mathbf{F}_{ac} \\
                            \mathbf{F}_{ba} & \mathbf{F}_{bb} & \mathbf{F}_{bc} \\
                            \mathbf{F}_{ca} & \mathbf{F}_{cb} & \mathbf{F}_{cc}
                        \end{pmatrix}, \ \ \ \
     \mathbf{G}(p, q) = \begin{pmatrix}
                            \mathbf{G}_{aa} & \mathbf{G}_{ab} & \mathbf{G}_{ac} \\
                            \mathbf{G}_{ba} & \mathbf{G}_{bb} & \mathbf{G}_{bc} \\
                            \mathbf{G}_{ca} & \mathbf{G}_{cb} & \mathbf{G}_{cc}
                        \end{pmatrix}.
 \end{align}
A non-zero contribution to $\mathbf{F}$ can only occur for two edges which are shared by a vertex. Two neighbouring edges which are shifted relatively by a single positive or negative step in the parameters $k,l$ are, in Fourier space, related by a factor $e^{\pm ip}$ and $e^{\pm iq}$, respectively. From the neighbourhood relations of the triangulation of the flat torus and these factors, we obtain the following components of $\mathbf{F}$,
 \begin{align}
     \mathbf{F}_{aa} &= 2 - e^{ip} - e^{-ip} = 4 \sin^2\left(\frac{p}{2}\right) \\
     \mathbf{F}_{bb} &= 2 - e^{i(p + q)} - e^{-i(p + q)} = 4 \sin^2\left(\frac{p + q}{2}\right) \\
     \mathbf{F}_{cc} &= 2 - e^{iq} - e^{-iq} = 4\sin^2\left(\frac{q}{2}\right) \\
     \mathbf{F}_{ab} &= 1 - e^{-ip} + e^{iq} - e^{i (p + q)} = 4 e^{iq/2} \sin\left(\frac{p}{2}\right)\sin\left(\frac{p + q}{2}\right) \\
     \mathbf{F}_{ac} &= 1 - e^{-ip} - e^{-iq} + e^{i (p - q)} = 4 e^{i(p - q)/2} \sin\left(\frac{p}{2}\right)\sin\left(\frac{q}{2}\right) \\
     \mathbf{F}_{bc} &= 1 + e^{-ip} - e^{iq} - e^{i(p + q)} = 4 e^{-ip/2}\sin\left(\frac{q}{2}\right)\sin\left(\frac{p + q}{2}\right) \\
     \mathbf{F}_{ba} &= \mathbf{F}^{*}_{ab}, \mathbf{F}_{ca} = \mathbf{F}^{*}_{ac}, \mathbf{F}_{cb} = \mathbf{F}^{*}_{bc}.
 \end{align}
where $F*$ is the conjugate transpose of $F$. We derive the matrix $\mathbf{G}$ as a function of $p$ and $q$ in a similar fashion. A non-zero contribution of $\mathbf{G}$ can only occur if two edges are shared by a triangle. Using the same factors as before, we find the following components of $\mathbf{G}$,
\begin{align}
    \mathbf{G}_{aa} &= \mathbf{G}_{bb} = \mathbf{G}_{cc} = 2\\
    \mathbf{G}_{ab} &= -1 - e^{iq} = -2 e^{iq/2} \cos\left(\frac{q}{2}\right) \\
    \mathbf{G}_{ac} &= e^{-ip} + e^{iq} = 2 e^{-i(p - q)/2} \cos\left(\frac{p + q}{2}\right) \\
    \mathbf{G}_{bc} &= -1 - e^{-ip} = -2 e^{-ip/2}\cos\left(\frac{p}{2}\right) \\
    \mathbf{G}_{ba} &= \mathbf{G}^{*}_{ab}, \mathbf{G}_{ca} = \mathbf{G}^{*}_{ac}, \mathbf{G}_{cb} = \mathbf{G}^{*}_{bc}.
\end{align}
The eigenvectors of $\mathbf{F}$ and $\mathbf{G}$ are orthogonal. The eigenvalues of $\Delta^{(1)}$ are therefore either an eigenvalue $\lambda^\mathbf{F}(p,q)$ of $\mathbf{F}$ or an eigenvalue $\lambda^\mathbf{G}(p,q)$ of $\mathbf{G}$,
\begin{align}
    \lambda_{1}^{\mathbf{F}}(p, q) &= 0 \\
    \lambda_{2}^{\mathbf{F}}(p, q) &= 0 \\
    \lambda_{3}^{\mathbf{F}}(p, q) &= 4\left(\sin^{2}\left(\frac{p}{2}\right) + \sin^{2}\left(\frac{p + q}{2}\right) + \sin^{2}\left(\frac{q}{2}\right)\right)
\end{align}
and 
\begin{align}
    \lambda_{1}^{\mathbf{G}}(p, q) &= 3 - \sqrt{9 - 4\left(\sin^{2}\left(\frac{p}{2}\right) + \sin^{2}\left(\frac{p + q}{2}\right) + \sin^{2}\left(\frac{q}{2}\right)\right)} \\
    \lambda_{2}^{\mathbf{G}}(p, q) &= 3 + \sqrt{9 - 4\left(\sin^{2}\left(\frac{p}{2}\right) + \sin^{2}\left(\frac{p + q}{2}\right) + \sin^{2}\left(\frac{q}{2}\right)\right)} \\
    \lambda_{3}^{\mathbf{G}}(p, q) &= 0.
\end{align}
The eigenvalues $\lambda^\mathbf{F}(p,q)$ and $\lambda^\mathbf{G}(p,q)$ are given as a function of $p$ and $q$. We note that $\lambda_i^\mathbf{F}(p,q)$ is equal to $0$ if $\lambda_i^\mathbf{G}(p,q)$ is not and vice versa. The eigenvalues $\lambda_i(p,q)$ of $\Delta^{(1)}$ can therefore be written as $\lambda_i(p,q) = \lambda_i^{\mathbf{F}}(p,q) + \lambda_i^{\mathbf{G}}(p,q)$, for $i \in \{1,...,3 \}$. Using Eq. (\ref{eq:d3.22}), we obtain the spectral dimension $D^{(1)}$ from the eigenvalues $\lambda_i(p,q)$. Fig. \ref{dfig:3proof-of-concept} shows the curve of $D^{(1)}$ as a function of $\tau$ derived from the analytical calculation of the spectrum of $\Delta^{(1)}$ for a triangulation of a flat two-dimensional torus of size $L_x\times L_y$.

\bibliographystyle{unsrt}
\bibliography{references}

\begin{thebibliography}{10}

\bibitem{Carlip_2017}
S~Carlip.
\newblock Dimension and dimensional reduction in quantum gravity.
\newblock {\em Classical and Quantum Gravity}, 34(19):193001, sep 2017.

\bibitem{Carlip}
Steven Carlip.
\newblock Dimension and dimensional reduction in quantum gravity.
\newblock {\em Universe}, 5(3), 2019.

\bibitem{Lifshitz}
Petr Ho\ifmmode~\check{r}\else \v{r}\fi{}ava.
\newblock Spectral dimension of the universe in quantum gravity at a lifshitz
  point.
\newblock {\em Phys. Rev. Lett.}, 102:161301, Apr 2009.

\bibitem{Trzesniewski}
Micha\l{} Eckstein and Tomasz Trze\ifmmode~\acute{s}\else \'{s}\fi{}niewski.
\newblock Spectral dimensions and dimension spectra of quantum spacetimes.
\newblock {\em Phys. Rev. D}, 102:086003, Oct 2020.

\bibitem{Calcagni1}
Gianluca Calcagni.
\newblock {\em Quantum Gravity and Gravitational-Wave Astronomy}, pages 1--27.
\newblock Springer Singapore, Singapore, 2020.

\bibitem{Calcagni2}
Gianluca Calcagni, Sachiko Kuroyanagi, Sylvain Marsat, Mairi Sakellariadou,
  Nicola Tamanini, and Gianmassimo Tasinato.
\newblock Gravitational-wave luminosity distance in quantum gravity.
\newblock {\em Physics Letters B}, 798:135000, 2019.

\bibitem{Calcagni3}
Gianluca Calcagni, Sachiko Kuroyanagi, Sylvain Marsat, Mairi Sakellariadou,
  Nicola Tamanini, and Gianmassimo Tasinato.
\newblock Quantum gravity and gravitational-wave astronomy, 2019.

\bibitem{Oriti}
Gianluca Calcagni, Daniele Oriti, and Johannes Thürigen.
\newblock Laplacians on discrete and quantum geometries.
\newblock {\em Classical and Quantum Gravity}, 30(12):125006, 2013.

\bibitem{SpecDimCDT1}
J.~Ambj\o{}rn, J.~Jurkiewicz, and R.~Loll.
\newblock The spectral dimension of the universe is scale dependent.
\newblock {\em Phys. Rev. Lett.}, 95:171301, Oct 2005.

\bibitem{CDTReview2012}
J.~Ambjørn, A.~Görlich, J.~Jurkiewicz, and R.~Loll.
\newblock Nonperturbative quantum gravity.
\newblock {\em Physics Reports}, 519(4):127--210, 2012.
\newblock Nonperturbative Quantum Gravity.

\bibitem{CDTReview2019}
R~Loll.
\newblock Quantum gravity from causal dynamical triangulations: a review.
\newblock {\em Classical and Quantum Gravity}, 37(1):013002, 2019.

\bibitem{CDTReview2021}
Jan Ambjorn, Zbigniew Drogosz, Jakub Gizbert-Studnicki, Andrzej Görlich, Jerzy
  Jurkiewicz, and Dániel Németh.
\newblock {CDT} quantum toroidal spacetimes: An overview.
\newblock {\em Universe}, 7(4), 2021.

\bibitem{ComLap}
J.~R. Munkres.
\newblock {\em Elements of Algebraic Topology}.
\newblock Addison-Wesley, 1984.

\bibitem{Regge:1961px}
T.~Regge.
\newblock {General Relativity Without Coordinates}.
\newblock {\em Nuovo Cim.}, 19:558--571, 1961.

\bibitem{thephasestruc}
J.~Ambjørn, J~Gizbert-Studnicki, J.~Jurkiewicz, A.~Görlich, and D~Németh.
\newblock The phase structure of causal dynamical triangulations with toroidal
  spatial topology.
\newblock {\em Journal of High Energy Physics}, page 111, 2018.

\bibitem{SpecDimCDT2}
J.~Ambj\o{}rn, J.~Jurkiewicz, and R.~Loll.
\newblock Reconstructing the universe.
\newblock {\em Phys. Rev. D}, 72:064014, 2005.

\bibitem{nonpertquantumdes}
J.~Ambj\o{}rn, A.~G\"orlich, J.~Jurkiewicz, and R.~Loll.
\newblock Nonperturbative quantum de {S}itter universe.
\newblock {\em Phys. Rev. D}, 78:063544, Sep 2008.

\bibitem{impacttop}
J.~Ambj\o{}rn, Z.~Drogosz, J.~Gizbert-Studnicki, A.~G\"orlich, J.~Jurkiewicz,
  and D.~Nemeth.
\newblock Impact of topology in causal dynamical triangulations quantum
  gravity.
\newblock {\em Phys. Rev. D}, 94:044010, 2016.

\bibitem{AMBJORN2010420}
J.~Ambjørn, A.~Görlich, J.~Jurkiewicz, and R.~Loll.
\newblock Geometry of the quantum universe.
\newblock {\em Physics Letters B}, 690(4):420--426, 2010.

\bibitem{Zbyszek2021}
J.~Ambjorn, Z.~Drogosz, A.~G\"orlich, and J.~Jurkiewicz.
\newblock Properties of dynamical fractal geometries in the model of causal
  dynamical triangulations.
\newblock {\em Phys. Rev. D}, 103:086022, 2021.

\bibitem{Craioveanu}
M.~(Mircea) Craioveanu, Mircea. Puta, and Themistocles~M. Rassias.
\newblock {\em {Old and new aspects in spectral geometry}}.
\newblock Springer, Dordrecht, 2001.

\bibitem{He}
Yue He.
\newblock A lower bound for the first eigenvalue in the laplacian operator on
  compact riemannian manifolds.
\newblock {\em Journal of Geometry and Physics}, 71:73--84, 2013.

\bibitem{MCA2}
Marcus Reitz and Ginestra Bianconi.
\newblock The higher-order spectrum of simplicial complexes: a renormalization
  group approach.
\newblock {\em Journal of Physics A: Mathematical and Theoretical},
  53(29):295001, 2020.

\bibitem{Desbrun}
Mathieu Desbrun, Anil~N. Hirani, Melvin Leok, and Jerrold~E. Marsden.
\newblock Discrete exterior calculus, 2005.

\bibitem{MCA1}
J~Brunekreef and M~Reitz.
\newblock Approximate killing symmetries in non-perturbative quantum gravity.
\newblock {\em Classical and Quantum Gravity}, 38(13):135009, 2021.

\bibitem{bossavit1998}
Alain Bossavit.
\newblock {\em Computational Electromagnetism: Variational Formulations,
  Complementarity, Edge Elements}.
\newblock Academic Press, 1998.

\bibitem{hirani_discrete_2003}
A.~Hirani.
\newblock {\em Discrete Exterior Calculus}.
\newblock {PhD} dissertation, Caltech, 2003.

\bibitem{Ben-Chen}
Mirela Ben-Chen, Adrian Butscher, Justin Solomon, and Leonidas Guibas.
\newblock On discrete killing vector fields and patterns on surfaces.
\newblock {\em Computer Graphics Forum}, 29(5):1701--1711, 2010.

\bibitem{topinduced}
J.~Ambjørn, J.~Gizbert-Studnicki, A.~Görlich, and D~Németh.
\newblock Topology induced first-order phase transitions in lattice quantum
  gravity.
\newblock {\em Journal of High Energy Physics}, page 103, 2022.

\bibitem{Ambjorn:2014gsa}
J~Ambj{\o}rn, A~Görlich, J~Jurkiewicz, A~Kreienbuehl, and R~Loll.
\newblock Renormalization group flow in {CDT}.
\newblock {\em Classical and Quantum Gravity}, 31(16):165003, jul 2014.

\bibitem{Renormalization_Ambjorn2020}
Jan Ambjorn, Jakub Gizbert-Studnicki, Andrzej Görlich, Jerzy Jurkiewicz, and
  Renate Loll.
\newblock Renormalization in quantum theories of geometry.
\newblock {\em Frontiers in Physics}, 8, 2020.

\bibitem{towardscontinum}
J.~Ambjorn, D.~N. Coumbe, J.~Gizbert-Studnicki, and J.~Jurkiewicz.
\newblock Searching for a continuum limit in causal dynamical triangulation
  quantum gravity.
\newblock {\em Phys. Rev. D}, 93:104032, May 2016.

\bibitem{cdt_desit_fi}
J.~Ambj\o{}rn, J.~Jurkiewicz, and R.~Loll.
\newblock Emergence of a 4d world from causal quantum gravity.
\newblock {\em Phys. Rev. Lett.}, 93:131301, Sep 2004.

\end{thebibliography}

\end{document}